\PassOptionsToPackage{table}{xcolor}
\documentclass[conference]{IEEEtran}
\IEEEoverridecommandlockouts
\usepackage{cite}
\usepackage{amsmath,amssymb,amsfonts}
\usepackage{textcomp}

\usepackage[utf8]{inputenc}
\usepackage[T1]{fontenc}
\usepackage{microtype}
\usepackage[a-2b]{pdfx}
\usepackage{graphicx}
\usepackage{balance}  %

\usepackage{subfigure}
\usepackage{multirow}
\usepackage{algorithm}
\usepackage{algorithmic}
\usepackage{color}
\usepackage{listings}
\usepackage{float}
\usepackage{xcolor}
\usepackage{outlines}
\usepackage[normalem]{ulem}
\usepackage{cleveref}
\usepackage{enumitem}
\usepackage{soul}
\usepackage{cuted, lipsum}
\usepackage[listings,skins,breakable]{tcolorbox}
\usepackage{booktabs}
\usepackage[table]{xcolor}
\usepackage{colortbl}
\usepackage{marginnote}

\usepackage{amsmath}
\usepackage{subscript}
\usepackage[skip=0pt]{caption}
\usepackage[labelfont=bf]{caption}
\usepackage{xcolor,colortbl}

\newcommand{\rightquadcirc}{\put(0,0){\oval(10,10)[tr]}\phantom{\circ}}

\newcommand{\eat}[1]{}
\def\BibTeX{{\rm B\kern-.05em{\sc i\kern-.025em b}\kern-.08em
    T\kern-.1667em\lower.7ex\hbox{E}\kern-.125emX}}
\begin{document}

\title{\LARGE{DeepMapping: Learned Data Mapping for Lossless Compression and Efficient Lookup}
}

\author{\IEEEauthorblockN{Lixi Zhou}
\IEEEauthorblockA{
\textit{Arizona State University}\\
Tempe, USA \\
lixi.zhou@asu.edu}
\and
\IEEEauthorblockN{ K. Sel\c{c}uk Candan}
\IEEEauthorblockA{
\textit{Arizona State University}\\
Tempe, USA \\
candan@asu.edu}
\and
\IEEEauthorblockN{ Jia Zou}
\IEEEauthorblockA{
\textit{Arizona State University}\\
Tempe, USA \\
jia.zou@asu.edu}
}

\maketitle

\begin{abstract}
Storing tabular data to balance storage and query efficiency is a long-standing research question in the database community. 
In this work, we argue and show that a novel {\em DeepMapping} abstraction, which relies on the impressive {\em memorization} capabilities of deep neural networks, can provide better storage cost, better latency, and better run-time memory footprint, all at the same time. Such unique properties may benefit a broad class of use cases in capacity-limited devices. 
Our proposed DeepMapping abstraction transforms a dataset into multiple key-value mappings and constructs a multi-tasking neural network model that outputs the corresponding \textit{values} for a given input \textit{key}.
To deal with memorization errors, DeepMapping couples the learned neural network with a lightweight auxiliary data structure capable of correcting mistakes. The auxiliary structure design further enables DeepMapping to efficiently deal with insertions, deletions, and updates even without retraining the mapping.  
We propose a multi-task search strategy for selecting the hybrid DeepMapping structures (including model architecture and auxiliary structure) with a desirable trade-off among memorization capacity, size, and efficiency. 
Extensive experiments with a real-world dataset, synthetic and benchmark datasets, including TPC-H and TPC-DS, 
demonstrated that the DeepMapping approach can better balance the retrieving speed and compression ratio against several cutting-edge competitors. 
\end{abstract}

\section{Introduction}
\label{sec:intro}
Real-time computations are increasingly pushed to edge servers with limited computational and storage capabilities for efficiency, cost, and privacy reasons. It is nevertheless critical to balance storage (e.g., on-disk, in-memory) and computational costs (e.g., query execution latency) on these platforms to achieve real-time response. 
Most existing works in balancing compression and retrieval tasks rely on two approaches: (1) using regression to approximate segmented numerical data, such as ModelarDB~\cite{jensen2018modelardb}, and (2) enforcing ordering over compression, such as segregated encoding~\cite{raman2006wring}. Unfortunately, both approaches provide \textit{sub-optimal} latency for exact-match queries (lookups): the former requires scanning each segment, while the latter requires binary search.
However, lookup is important for many emerging edge applications. Taking self-serve retailing~\cite{wang2021does, kim2023self}, as an example, to provide robust service in unreliable environments, the edge device should be able to manage transaction and inventory data locally for random lookups and updates. Other examples include large-scale manufacturing, where edge devices must store product quality and defect categories for looking up the quality of individual products and updating categorical quality control parameters~\cite{xu2019big}, and autonomous robots lookup information (e.g., brands and prices) about objects nearby from a local database~\cite{ye2021hierarchical}. 
Existing solutions are not sufficiently effective in integrating compression and indexing techniques to achieve low storage costs and low query latency simultaneously: (de)compression operations are usually computational-intensive, whereas indexing techniques impose additional storage overheads. 

\begin{figure}[t]
\centering{%
   \includegraphics[width=3in]{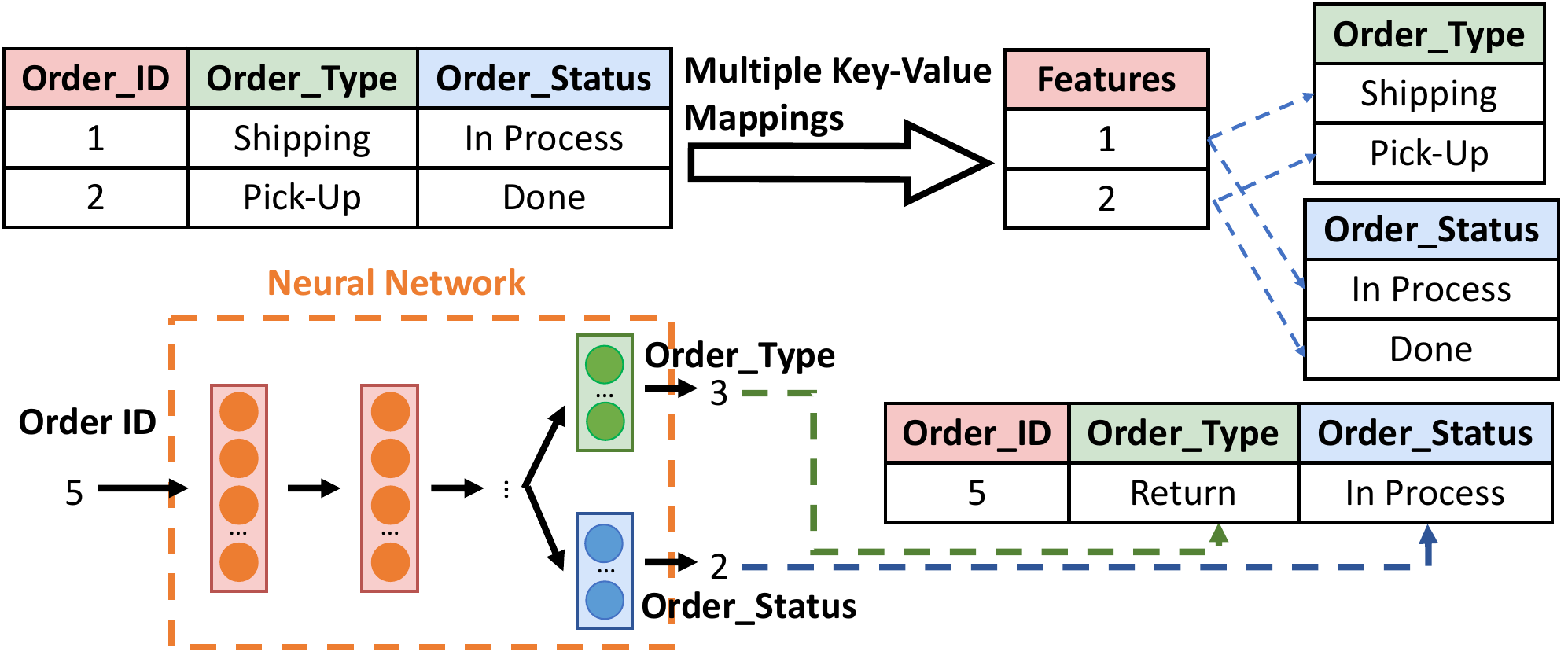}  
}
\caption{\label{fig:nn-comp-data} \small DeepMapping relies on neural networks to memorize key-value mapping in tabular data.}
\end{figure}

\vspace{3pt}
To address these problems, in this work, we argue for a novel data abstraction, called {\em Deep Learned Data Mapping (or DeepMapping)}, which leverages deep neural networks to integrate the compression and indexing capabilities seamlessly. 
The main idea is to leverage the impressive learning capabilities of neural networks~\cite{DBLP:conf/cidr/Zou21} for compression and efficient query processing of \textit{key-value based maps}.
As illustrated in Figure~\ref{fig:nn-comp-data}, a tabular \texttt{Orders} dataset is represented as two mappings from the key \texttt{Order\_ID} to the attributes \texttt{Order\_Type} and \texttt{Order\_Status}, respectively. In DeepMapping, these two mappings are stored as one multi-tasking neural network which takes \texttt{Order\_ID} as input feature and outputs \texttt{Order\_Type} and \texttt{Order\_Status} as labels. DeepMapping is motivated by \underline{opportunities} brought by deep neural network models:

\noindent
$\bullet$ \textbf{Compressibility opportunities.} 
Compressibility is a function of the statistical properties of the data, such as the underlying key-value correlations. If structures or patterns exist in the underlying datasets, deep learning models often have significantly smaller sizes than their training datasets.  For example, the common crawl dataset is $220$ Terabytes in size~\cite{patel2020introduction}, yet the language-agnostic BERT sentence embedding model trained on the dataset is just $1.63$ Gigabytes in size~\cite{tfhub, conneau2019unsupervised}.

\noindent
$\bullet$
\textbf{Hardware acceleration opportunities}. In general, (batched) inference computations of a neural network can be accelerated using hardware such as GPU processors. %
ow-end GPUs that cost hundreds of dollars, equipped with $4-16$ GB memory, are becoming widely available at the edge~\cite{edge-gpu}.

While there exist related works exploiting these opportunities for compression and learned indexing,  respectively (Section~\ref{sec:relworks}), DeepMapping is facing \textit{unique} \underline{challenges}:

\noindent
$\bullet$ 
 \textbf{The accuracy challenge.} 
 Unlike numerical data for which accuracy loss caused by compression is acceptable, categorical data usually requires lossless compression with $\textbf{100\%}$ accuracy. Although the universal approximation theorem~\cite{cybenko1989approximation, hornik1989multilayer, huang2003learning} states that given a continuous function defined on a certain domain, even a neural network with a single hidden layer can approximate this continuous function to arbitrary precision, the resulting layer's size could be significantly larger than the dataset size. In addition, it is challenging for a deep neural network to recognize non-existing keys and avoid the data existence hallucination, as detailed in Sec.~\ref{sec:accuracy-assure}. %
 
\vspace{3pt}
\noindent
$\bullet$  \textbf{The modification challenge.} Although simple lookup queries, such as \texttt{\textbf{SELECT} Order\_Type \textbf{FROM} Orders \textbf{WHERE} Order\_ID=19}, can be implemented as inference operations over neural models, it is not straightforward to implement \texttt{update}, \texttt{insertion}, \texttt{deletion}. Particularly, 
relying on the incremental model (un)learning for the above operations 
can result in "catastrophic forgetting" issues~\cite{mccloskey1989catastrophic}. %

\vspace{3pt}
\noindent
$\bullet$ \textbf{The model search challenge.} The overall performance (i.e., size reduction and accuracy) of DeepMapping depends on the underlying neural network model. Obviously, it would be difficult and laborious for developers to search for architectures to replace tabular data manually.

\begin{figure*}[!h] 
\centering
\includegraphics[width=0.9\textwidth]{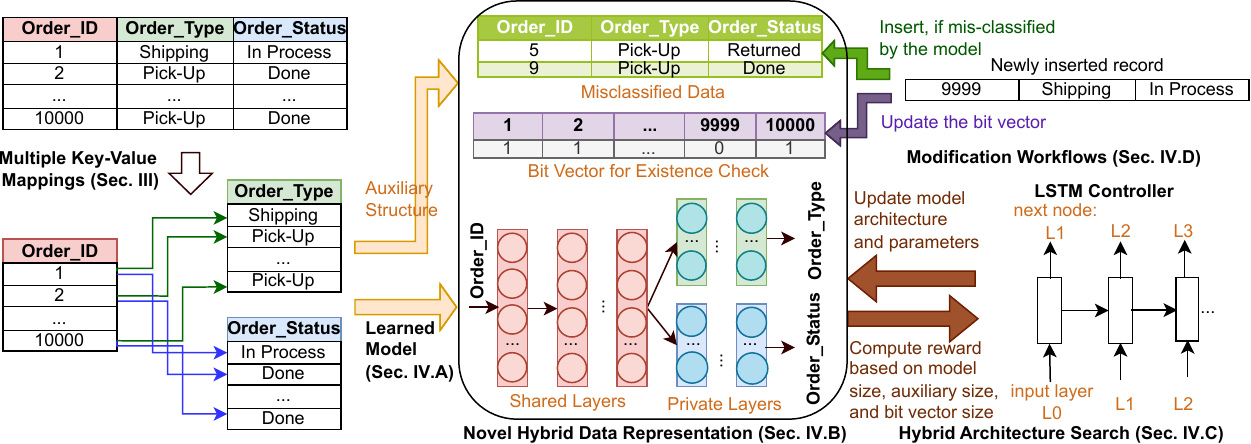}
\caption[]{\label{fig:overview} \small
Overview of the proposed neural network-based data compression methods.
}
\vspace{-15pt}
\end{figure*}

\vspace{5pt}
To address these challenges, we introduce \textbf{\textit{a novel DeepMaping framework}},  outlined in Figure~\ref{fig:overview} (and detailed in Section~\ref{sec:deepmapping}), with the following  \underline{\textbf{key contributions}}:

\vspace{3pt}
\noindent
{\textbf{(1) A Novel Hybrid Data Representation (Sec.~\ref{sec:accuracy-assure})}} 
Instead of trying to increase the model size to achieve the last-mile accuracy, we propose to couple a relatively simple and imperfect neural network with a lightweight auxiliary structure that manages the misclassified data and an existence indexing:

\noindent
\underline{\textit{
$\bullet$ 
A Compact, Multi-Task Neural Network Model}} is trained to capture the correlations between the key (i.e., input features) 
and the values (i.e., labels) of a given 
key-value mapping.
To memorize the data with multiple attributes, we propose to train a multi-task neural network, where each output layer outputs the value for its corresponding attribute.
The query answering process takes a (batch of) query key(s) as the input and outputs the predicted value(s).

\noindent
\underline{\textit{
$\bullet$ 
An Auxiliary Accuracy Assurance Structure}} compresses the mappings that are \textit{misclassified} by the model and a snapshot of the keys, to ensure query accuracy. 
While the neural network memorizes a significant portion of the data, the auxiliary data structure memorizes misclassified data to achieve $100\%$ overall accuracy on data query;  
An additional bit vector is used to record the existence of the data. %

\vspace{5pt}
\noindent
{\textbf{(2) 
Multi-Task Hybrid Architecture Search (MHAS)
(Sec.~\ref{sec:model-search})}}
The objective of the model search process is to minimize the overall size of the hybrid architecture: (a) 
The architecture should maximize the sharing of model layers/parameters across the inference tasks corresponding to different mappings. (b) The 
model should specialize well for tables with attributes that have heterogeneous types and distributions.
We abstract the search space as a directed acyclic graph (DAG), comprising a collection of $k$ nodes, where each node represents a \textit{configurable} layer and each edge represents a data flow from the source node to the target node. A candidate model is sampled as a sub-graph from the DAG. We further propose a multi-task search strategy based on deep reinforcement learning that adaptively tunes the number of shared and private layers and the sizes of the layers.%

\vspace{5pt}
\noindent
{\textbf{(3) Workflows for Insertions, Deletions, and Updates (Sec.~\ref{sec:data-manipulation})}} 
To address the challenge of supporting data modifications, we propose a lazy-update process that re-purposes the lightweight auxiliary structure outlined above by \textit{materializing the modification operations in this structure} if the model cannot capture those modifications. The system triggers retraining of the neural network model only when the size of the auxiliary structure exceeds a threshold.

\eat{
\vspace{5pt}
\noindent
{\textbf{Other Data Operations.} While, in this paper, we focus on mapping and data updates, we note that the proposed data structure can also support more complex operations.
For example, \texttt{scan}s can be implemented efficiently by running a batch inference over the samples in a given key range. Then, the existence index can be used to prune the inference results. We leave investigating more complex data operations and query processing as future work.
}
}

\vspace{5pt} 
We implemented the proposed approach and conducted extensive experiments on TPC-H, TPC-DS, synthetic datasets, and a real-world crop dataset. We considered baselines, such as hash-based and array-based representations of the datasets partitioned and compressed by cutting-edge compression techniques.%
{The evaluation results demonstrated that the DeepMapping approach better balances efficiency, offline and run-time storage requirements against the baselines in our target scenarios. Especially in scenarios with limited memory capacity, DeepMapping achieved up to $\textbf{15}\times$ speedup by alleviating the I/O and decompression costs. In such cases, DeepMapping runs in memory. However, the baselines require loading (evicted) partitions (back) into the memory and decompressing these to answer random queries.}

\eat{
\begin{figure}[t]
\centering
\includegraphics[width=0.49\textwidth]{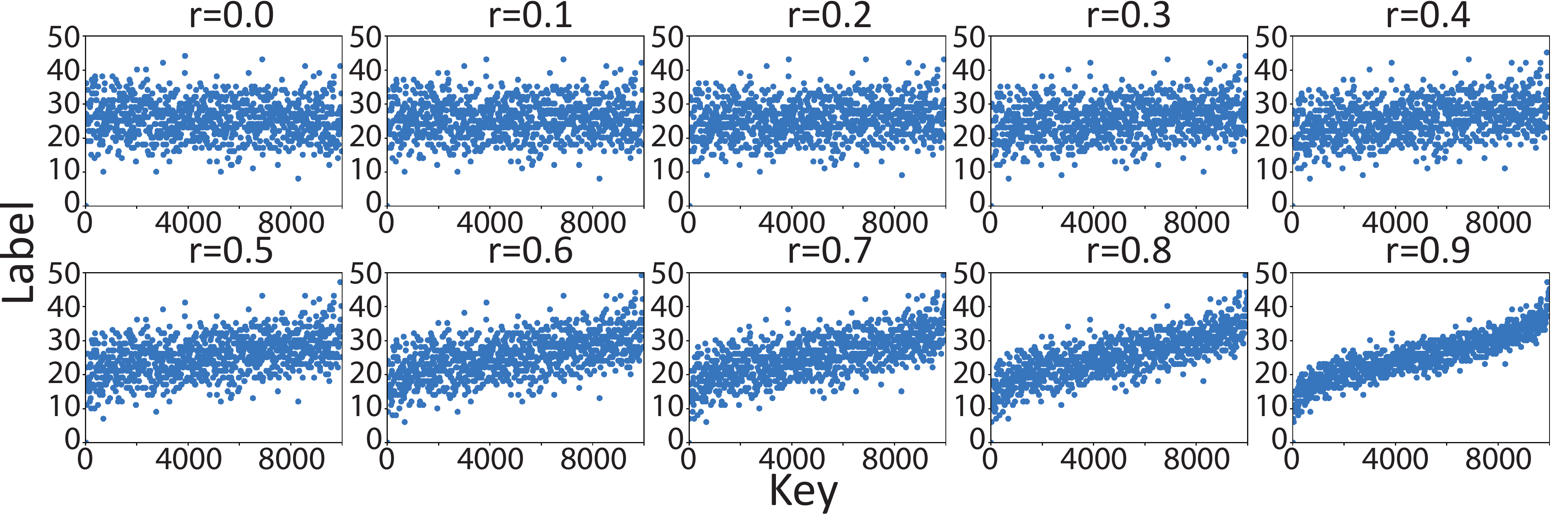}
\centerline{(a)  Data with low, $r=0.0$, to  high, $r= 0.9$, correlation
}
 \includegraphics[width=3.3in]{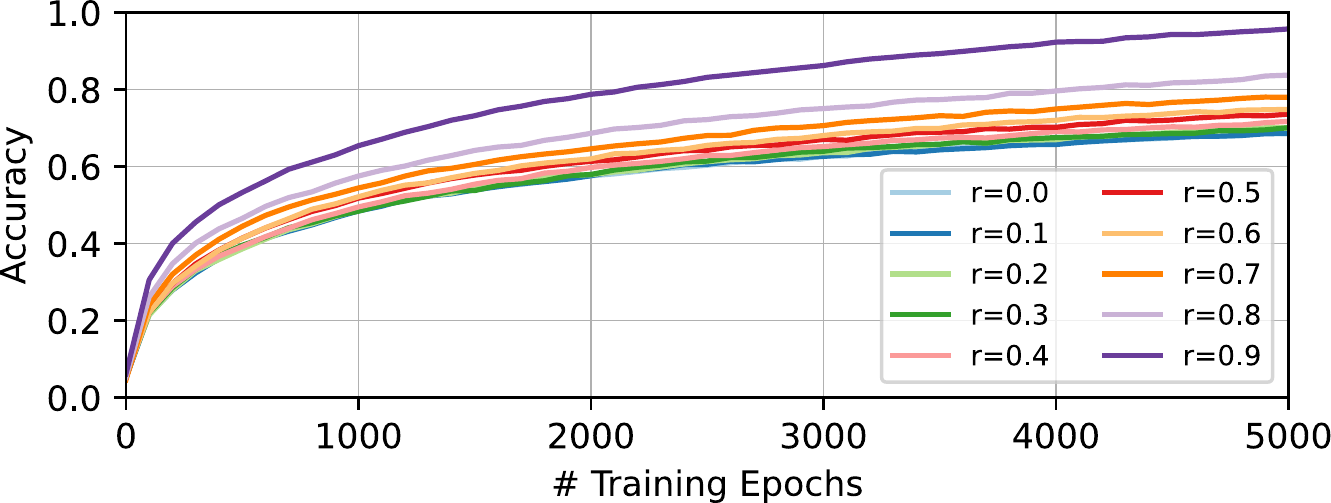} 
 \centerline{(b) Learning capacity (training epochs vs. accuracy)} 
\vspace*{0.05in}
\caption{\label{fig:synthetic-overview} \label{fig:synthetic_data_train_acc}
\small
 Learning capacity of a two-layer fully-connected neural network for data with different statistical properties.
}
\end{figure}
}

\vspace{-5pt}
\section{Related Works}\label{sec:relworks}
\noindent{\bf Learned Indexing.} 
Driven by the desirable properties of the neural networks outlined in Section~\ref{sec:intro}, in recent years,  {\em learned index} structures\cite{kraska2018case, ferragina2020pgm, kipf2020radixspline, ding2020tsunami, ding2020alex, li2020lisa, nathan2020learning, pandey2020case, qi2020effectively}, have been proposed to improve the computational efficiency of indexing structures.
These techniques apply machine learning to capture the correlation between the keys and the positions of the queried values in underlying storage. There also exist works~\cite{sabek2022can, sabek2021case, hentschel2022entropy, sabek2021learned} that use a machine learning model to replace the hash function for hash indexing. %
The problem we target in this paper is different from learned indexes in several critical ways:

\begin{itemize}[leftmargin=*]
\item Generally speaking, learned indexing predicts positions in a (sorted) array, which is commonly posed as a regression task. We, however, are aiming to learn direct key-value mapping  -- moreover, the values can be discrete or categorical.

\item For learned indexing, if the queried value is not found in the predicted position, the search can continue on the (sorted) array until the value is found or determined as non-existing. This is not possible for learning a direct key-value mapping.  

\item 
Learned indexing only compresses the indexing structure, but will not compress the data. Our DeepMapping method combines both losslesscompression and indexing and strikes an even better trade-off between storage and lookup latency.
\end{itemize}

\noindent {\bf Compression.}
Semantic compression~\cite{jagadish1999semantic, babu2001spartan, jagadish2004itcompress, ilkhechi2020deepsqueeze} leverage correlations among columns, which could be captured by tree-based models, auto-encoding methods, and so on,  to store a minimum number of columns for restoring entire tuples with error bounded. However, these approaches cannot eliminate the errors. In addition, these approaches cannot avoid the decompression overheads and accelerate the lookups. 
Abundant syntactic compression works~\cite{kingma2019bit, yang2020improving, mentzer2020learning, jensen2018modelardb, huang2022compressing, liang2022sz3} use machine learning for error-bounded compression of high-dimensional data by levaraging a statistical model to closely capture the underlying distribution of the input data. However, these works mainly focus on numerical data, and their approaches are not extensible to lossless compression for categorical data. 
Several syntactic compression works also aim to balance the compression ratio and query speed. For example, ModelarDB~\cite{jensen2018modelardb} focuses on the compression and aggregation queries over streaming numerical time-series data by modeling data as multiple linear segments. Their work cannot be applied to the compression and lookups of categorical data. Segregated encoding~\cite{raman2006wring} allows query processing over compressed data without decompression overheads. However, the lookup queries require searching in ordered segregates, which is sub-optimal. %

\eat{
\noindent
\textbf{Approximate Query Processing (AQP).} Recent works applied learning-based techniques to improve AQP ~\cite{hilprecht2019deepdb, ma2019dbest, zeighami2021neurodb, DBLP:journals/pvldb/ZeighamiAGS22}. Their solution and theoretical bounds only work for range aggregation queries and do not consider compression as an objective. %
}

\section{Problem and Desiderata}
\label{sec:problem}

We first formulate the problem and desiderata as follows: 

\begin{itemize}[leftmargin=*]

\item \noindent{\bf Single-Relation, Single-Key Mapping.} %
Let $R(\underline{K_1}, \ldots, \underline{K_l}, V_1, \ldots, V_m)$ be a relation where ${\tt K}=(K_1, \ldots, K_l)$ defines a key that consists of $l$ attributes and $V_1$ through $V_m$ are $m$ value attributes. 
The goal is to identify a mapping data structure, $d_\mu()$  which, given a key ${\bf{k}}=(k_1, \ldots, k_l) \in domain({\tt K})$, and a target attribute $V_i \in \{V_1, \ldots, V_m\}$, it
returns 
$d_\mu({\bf{k}}, V_i) = 
\pi_{V_i}(\sigma_{K_1=k_1 \wedge \ldots \wedge K_l=k_l}(R))
$. 
Different from the key in relational models, a key in DeepMapping does not need to be a unique identifier. A key can consist of any attribute.  
While we focus on this problem, our approach is extensible to the following two problems (See~\cite{zhou2023deepmapping} for detailed formalization).

\item \noindent{\bf Single-Relation, Multiple-Key Mapping.}   
Note that in practice, the workload may look up values using different key columns from the same relation, thus requiring multiple mappings with different keys. \eat{We further represent the collection of keys as $\mathcal{K=({\tt K^1}, \ldots, {\tt K^s}})$. Therefore, we more generally seek a multiple-mapping data structure, $d_\mu()$, which takes a key ${\bf{k}}^i=(k_1, \ldots, k_{l}) \in domain({\tt K^i} \in \mathcal{K})$
along with a target value attribute $V_j \in \{V_1, \ldots, V_m\}$, 
it returns $d_\mu({\bf{k}}^i, V_j) = 
\pi_{V_j}(\sigma_{K^i_{1}=k_1 \wedge \ldots \wedge K^i_{l}=k_l}(R))
$.}

\item \noindent{\bf Multiple-Relation, Multiple-Key Mapping.}
Furthermore, in many contexts, such as databases with star schemas, the same attribute from one relation (e.g., the fact table) may reference attributes in other relations (e.g., the dimension tables). Cross-table lookups require multiple-relation multiple-key mapping. \eat{Let us consider $\mathcal{R} = \{R_1, \ldots, R_r\}$ be a set of relations. In this case, we seek a multiple-mapping data structure, $d_\mu()$, which takes a key ${\bf{k}}^i=(k_1, \ldots, k_{l}) \in domain({\tt K^i} \in \mathcal{K})$,  a target value attribute $V_j \in \{V_1, \ldots, V_m\}$,
along with a target relation $R_u \in \mathcal{R}$,
and it returns $d_\mu({\bf{k}}^i, V_j, R_u) = 
\pi_{V_j}(\sigma_{K^i_{1}=k_1 \wedge \ldots \wedge K^i_{l}=k_l}(R_u))
$.}
\end{itemize}

\noindent For any of these three alternative problem scenarios, our key desiderata from the mapping data structure, $d_\mu()$,  are as follows:
\begin{itemize}[leftmargin=*]
    \item {\bf Desideratum $\#$1 -Accuracy:} $d_\mu()$ is accurate -- i.e., it does not miss any data and it does not return any spurious results.
     \item {\bf Desideratum $\#$2 -Compressibility:} $d_\mu()$ structure is compact in offline disk storage and runtime memory footprint.
    \item {\bf Desideratum $\#$3 -Low latency:} The data structure $d_\mu()$ is efficient and, thus, provides low data retrieval latency.
    \item {\bf Desideratum $\#$4 -Updateability:} $d_\mu()$ is updateable with insertions of new key-value rows and deletions of some existing keys from the database. Moreover, the data structure also allows changing the value of an existing key.
    
\end{itemize}

{
\section{DeepMapping Architecture}
 \label{sec:deepmapping}

\subsection{Shared Multi-Task Network }
\label{sec:network}
This paper uses DeepMapping to manage key-value mappings, where the key and values are discrete (e.g., integers, strings, and categorical values). Without loss of generality, we consider a sequence of fully connected layers as the underlying neural network architecture, where the strings or categorical data
are encoded as integers using one-hot encoding before training and inference. %

To achieve high compression rates, some of these layers can be shared across multiple inference tasks within a relation and across relations that have foreign key reference relationships. At the same time,
other layers are private to each inference task to improve the accuracy of that particular inference task.  
\noindent
\textbf{How to divide shared/private layers?}  %
\eat{
}
 The first few layers of the neural network, which capture the structures common to both inference tasks, are shared. However, the latter layers of the neural network, which capture the output attribute-specific structures, are private to specialize the network for each output attribute. An example can be found in Figure ~\ref{fig:nn-comp-data}, which consists of two shared fully connected layers (in orange) and two private output layers, one for the $\tt Order\_Type$ column (in green) and the other for the $\tt Order\_Status$ column (in blue).
The number, types, and sizes of the shared layers and private layers are all determined by our multi-task hybrid architecture search (MHAS) algorithm, as detailed in Sec.~\ref{sec:model-search}. 
}

\subsection{Ensuring $\mathbf{100\%}$ Accuracy (Desideratum $\mathbf{\#1}$)} \label{sec:accuracy-assure}
DeepMapping should not miss any data and should not return spurious results. It is a challenging task because seeking an arbitrarily large model~\cite{cybenko1989approximation, hornik1989multilayer} to achieve $100\%$ accuracy is not appropriate, considering the desiderata include update ability, compressibility, and low latency. 
\eat{
\noindent{\bf Why not simply try to overfit the data?} By the universal approximation theorem~\cite{cybenko1989approximation, hornik1989multilayer}, a sufficiently large network should perfectly memorize the given data. However, such {\em overfitting} approach is undesirable for our problem: 
\begin{itemize}[leftmargin=*]
\item An overfitting model is often unnecessarily large (hence expensive to train and infer). To achieve the last-mile accuracy, e.g., to improve the accuracy from $90\%$ to $100\%$, the required neural network model size will significantly increase, often multiple times~\cite{kraska2018case}.
\item Models that are overfitting often do not generalize to unseen data, which makes them ineffective in generalizing to the prediction tasks for insertions.
\end{itemize}
}
{
In addition, when querying the non-existing data, it is challenging for the neural networks to tell whether the tuple exists accurately.
That is because the inference task will {\em predict} an output even when the data is not seen in the training data -- this makes the neural network-based solutions useful for generalizing. Still, in our context, any such output will be spurious and need to be avoided, as a {\em hallucination}.

\vspace*{3pt}
\noindent{\bf Solution: Lightweight Auxiliary Accuracy Assurance Structures. }
To address these issues, we propose to use a novel \textit{hybrid data representation}, consisting of (1) a compact \underline{\em neural network model} that memorizes the 
data (denoted as $M$), (2) an \underline{\em auxiliary accuracy assurance} table  (denoted as $T_{aux}$) that compresses any misclassified data, and (3) an \underline{\em existence bit vector} (denoted as $V_{exist}$), whose range corresponds to the key range -- intuitively, each bit marks the existence of the corresponding key. 

\subsubsection{The Auxiliary Structures ($T_{aux}$)}
To construct $T_{aux}$, the system runs all the keys in the input key-value mapping through the trained model and checks whether the inferred result matches the corresponding value. If not (i.e., the key-value mapping is misclassified), the key-value pair is stored in the auxiliary table. 
The misclassified key-value pairs are sorted by the key and equally partitioned. All misclassified key-value pairs in the same partition are sorted by key. (We NEVER rekey to make key ordering consistent with value orderings in either the model training process or the auxiliary data management). To further reduce the storage overhead, we apply Z-Standard or LZMA compression to each partition before they are stored.
In this work, we focus on keys that consist of discrete values and use a single dynamic bit array to serve an existence index for the keys, denoted as $V_{exist}$. Additionally, the decoding map (denoted as $f_{decode}$) that converts predicted labels from integer codes (resulting from one-hot encoding) to their original format, is also part of the auxiliary structure.

\begin{algorithm}[t]
\caption{(Parallel) Batch Key Lookup by DeepMapping}
\label{alg:lookup}
\small
\begin{algorithmic}[1]
\STATE INPUT: \\
$Q$: A vector of $k$ encoded query keys. \\
$M$: Pre-trained neural network. \\
$T_{aux}$: Auxiliary table that stores the misclassified data. \\
$V_{exist}$: Bit vector for existence check.\\
$f_{decode}$: Decoding map.
\STATE OUTPUT: $R$: A vector of $n$ queried values.

\STATE $R = M.infer(Q)$ // running in GPU as batch inference
\FOR{$i=1 \text{ to } n$}
    \IF{$V_{exist}[Q[i]] == 1$}
        \IF{$Q[i] \text{ exists in } T_{aux}$}
            \STATE $R[i] = T_{aux}[Q[i]]$
        \ENDIF
    \ELSE
        \STATE $R[i] = NULL$
    \ENDIF
\ENDFOR
\STATE $R = f_{decode}(R)$ // decoding
\RETURN $R$
\end{algorithmic}
\end{algorithm}

\subsubsection{Lookup Process}
The lookup process using the proposed hybrid data representation 
is illustrated in Algorithm ~\ref{alg:lookup} - the key aspects of the process are outlined below:

\noindent {\bf Inference.} DeepMapping leverages the ONNX runtime~\cite{onnxruntime} for optimizing the inference performance in edge environments.

\noindent {\bf Existence check.} 
The algorithm checks the existence of each query key in $V_{exist}$  to eliminate any spurious results.

\noindent {\bf Validation.} 
For any key that passes the existence check, the algorithm checks whether this key was misclassified by the model using the auxiliary table, $T_{aux}$. 
To look up the query key in the auxiliary table, the algorithm first locates its partition, brings it to the main memory if needed, and decompresses it; it then applies a binary search to look up the value within the partition. If the query key is located in the auxiliary table (i.e., the key-value pair was misclassified by the model), we return the value from the auxiliary table.
Otherwise, the model's output is returned as the result.
{
In memory-constrained devices, we free up the space of the least recently used (LRU) partition before loading the subsequent partition of the auxiliary table when the memory becomes insufficient. In this case, query keys in a batch are sorted before validation so that each partition is decompressed only once for each query batch.
}

Note that while their primary benefit is to ensure $100\%$ accuracy without necessitating a prohibitively large model, these auxiliary structures also enable \texttt{insert}/\texttt{update}/\texttt{delete} on the data. We provide details of these operations in Sec.~\ref{sec:data-manipulation}.
}

\subsection{Multi-Task Hybrid Architecture Search (Desiderata $\mathbf{\#2, 3}$)}
\label{sec:model-search}

A critical part of the DeepMapping initialization process is to select a neural network architecture to achieve the desiderata listed in Section~\ref{sec:problem}. 
Given a dataset $R$ that consists of $n$ tuples, where each tuple is represented as $<\textbf{x}, \textbf{y}>$. Here, $\textbf{x}$ represents a collection of one or more attributes that serve as the key so that each key uniquely and minimally identifies a tuple. $\textbf{y}$ represents a collection of $m$ attributes that are disjoint with $\textbf{x}$ and serve as the values. Our goal is to identify a
hybrid data representation, $\hat{M} = \langle M, T_{aux}, V_{exist}, f_{decode}\rangle$, consisting of a neural network model, $M$ along with the auxiliary structures, $T_{aux}$, $V_{exist}$, and $f_{decode}$, satisfying our desiderata.
As aforementioned in Sec.~\ref{sec:network}, we consider a multi-layer fully connected neural network to memorize a significant portion of the data in the given relation(s). 
As depicted in Figure~\ref{fig:nn-comp-data},  some layers (which help abstract the key) are shared across multiple data columns, while others are private to each output attribute.
Therefore, the problem is posed as a multi-task model search problem.

Neural architecture search is an active research area~\cite{li2020random, real2019regularized, bergstra2011algorithms, zoph2016neural, pham2018efficient, liu2018darts, guo2020single}. To develop our multi-task hybrid architecture search (MHAS) strategy, we build on the {\em efficient neural architecture search} (ENAS) strategy~\cite{pham2018efficient}, which 
supports parameter sharing across sampled model architectures.
While the original motivation of this parameter sharing is to improve the efficiency of NAS by forcing all sampled child models to share weights to eschew training each  model from scratch, in this paper, we argue that this approach can also
help reduce the model search overhead in multi-task learning by
encouraging parameter sharing across multiple tasks.
Our MHAS search algorithm, differs from ENAS in several ways: (a) MHAS is extended with the ability to search for multi-task models with shareable and private layers.
(b) %
To balance the compressibility brought by shared layers and the accuracy obtained by the private layers, MHAS searches within a search space with flexible numbers and sizes of hidden layers among each task's shared and private layers.
(c) Since our goal is not to search for a neural network but a hybrid structure, an objective function  (see Eq.~\ref{eq:loss}) governs the search that captures our desiderata.

MHAS consists of two components: 
{\em a search space}, which defines how the components of the underlying neural network can be connected to enumerate new neural architectures and 
{\em a controller algorithm} which seeks a structure for the target model within this search space. We describe both as follows:

\begin{figure}
    \centering
    \subfigure[Multi-task tree of DAGs]{\includegraphics[height=1.2in]{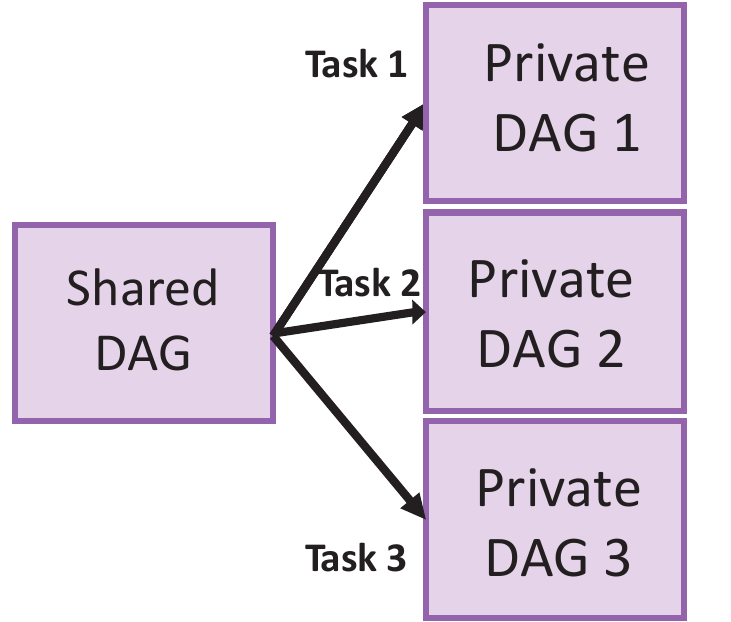}\hspace*{0.15in}\label{fig:nas-search-space-a}} 
    \subfigure[DAG with $\leq 2$ hidden layers]{\includegraphics[height=1.2in]{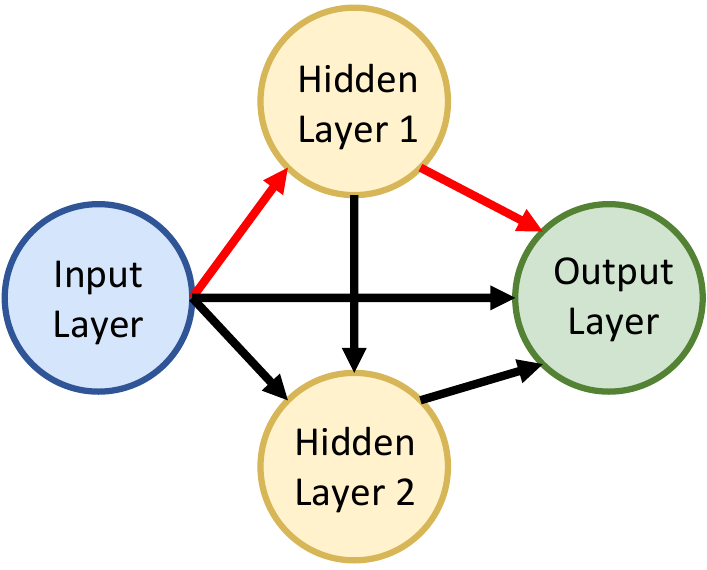}\label{fig:nas-search-space-b}} 
    \caption{\small (a) A high-level view of a candidate model and (b) a DAG (including all nodes and edges) represents the search space of one tree node in (a) -- here, the subgraph connected with the red edges illustrates a sampled network.
    }
    \label{fig:nas-search-space}
\end{figure}

\subsubsection{MHAS Multi-Task Search Space} \label{subsec:search-space}

Each model in the search space can be represented as a tree, where each non-leaf node represents a sequence of shared layers, and each leaf represents a sequence of private layers for a target column. 
Figure~\ref{fig:nas-search-space-a}  illustrates a class of models for memorizing a table with a key column and three non-key columns. It has four nodes: one for the shared layers (capturing the characteristics of the key) and three for the private layers, each corresponding to a target column.

Each node in the tree corresponds to a direct-acyclic graph (DAG). A DAG contains a node representing the input layer, a node representing the output layer, and multiple nodes representing candidate intermediate layers. Validated through our experiments, tabular data can be represented well using fully connected layers~\cite{liu2021survey}. Therefore, we consider each DAG node to be a fully-connected layer (with hyper-parameters, such as the number of neurons at each hidden layer also being searched). 
The DAG in Figure~\ref{fig:nas-search-space-b} represents a search space that contains up to two hidden layers.
Edges of the DAG represent all possible data flows among these layers (with each edge corresponding to a specific model parameter tensor that connects the two layers).
Given this, the model search process will enumerate 
subgraphs of the DAGs by {\em activating} and {\em deactivating} network edges -- each {\em activated}  directed edge represents a connection between two neural network layers. Deactivated edges represent connections that do not exist in that particular network. In Figure~\ref{fig:nas-search-space-b}, red color is used to highlight activated edges. 

Once an edge is activated and the destination node is a hidden layer, the controller will further choose the number of neurons for that layer  (i.e., layer size) flexibly to fully explore the entire search space.    
Suppose $N$ is the number of different layer sizes, $M$ is the maximum number of shared/private layers. The overall search space size is $
N^{2M}[\prod_{i=1}^{M}i \prod_{j=1}^{M}(2*j-1)]$
(i.e., $O(N^{2M} M! (2M-1)!!)$). The first term represents all possible layer size selections, while the second term represents all possible numbers of DAGs (i.e., layer connections).

\begin{algorithm}[t]
\caption{MHAS Model Search Algorithm}
\label{alg:model_search}
\small
\begin{algorithmic}[1]
\STATE INPUT:\\
$D$: A dataset to be memorized\\ 
$\theta$: The controller model parameters\\ 
$W$: The weights of all candidate layers in the search space\\
$N_t$, $N_m$, $N_c$: The  number of searching, model training, and controller training iterations\\
$D_{batch}$: batch size used in each model/controller training iteration\\
$m\_epochs$: number of epochs in each model training iteration
\STATE OUTPUT: optimal model structure $M$

\FOR{$i=1 \text{ to } N_t$}
    \STATE {\it \# Controller parameter $\theta$ is fixed}
    \IF{$i \ mod \ \lfloor \frac{N_t}{N_m} \rfloor == 0$}
        \STATE {\it \# Train a sampled model for $N_m$ iterations}
        \STATE Sample a model $M$ from the search space 
        \FOR{$k=1 \text{ to } m\_epochs$}
            
             \FOR{$D_{batch} \in D$}
                \STATE Update $W$ through $M.fit(D_{batch})$
             \ENDFOR
        \ENDFOR
    \ENDIF
    \IF{$i \ mod \ \lfloor \frac{N_t}{N_c} \rfloor == 0$}
        \STATE {\it \# Model weight $W$ is fixed}
        \FOR{$D_{batch} \in D$}
            \STATE Sample a model $M$ from the search space
            \STATE Update $\theta$ through $\mathcal{L}(M, D)$
        \ENDFOR
    \ENDIF
\ENDFOR
\RETURN $M$
\end{algorithmic}
\end{algorithm}

{ 

\subsubsection{Multi-Task Model Search Controller}\label{subsec:search-algo}

The DeepMapping multi-task model search algorithm is outlined in Algorithm ~\ref{alg:model_search}.
Since the goal is to learn a sequence of nodes (and train the model parameters at each edge), as in ENAS~\cite{pham2018efficient}, the DeepMapping controller is constructed as a long short-term memory (LSTM) neural network architecture.
The LSTM architecture samples decisions via softmax classifiers in an autoregressive fashion to derive a sequence of nodes in the DAG.
The
algorithm runs over $N_t$ iterations -- at each iteration, the algorithm 
alternatively 
trains the controller parameter $\theta$ in \textit{a controller training iteration} or  the weights of the sampled neural architecture $M$ through \textit{a model training iteration}:

\noindent
$\bullet$ During {\em a controller training iteration}, %
for each batch of the data, the controller samples a model from the search space and updates the controller parameter, $\theta$, through the loss function, $\mathcal{L}(\hat{M}, D)$ ($D$ represents the dataset to be compressed)%
:
\vspace{-1pt}
\begin{equation}
\label{eq:loss}
\frac{\text{size}(M) + \text{size}(T_{aux}) + \text{size}(V_{exist}) +\text{size}(f_{decode})}{size(D)}
\end{equation}

The controller samples a model by taking a node in the DAG as input and picking and \textit{configuring} the next node among the ones that connect to it; the selected node serves as the input for the next iteration of the process. This process repeats until the output node is selected.%

\noindent
$\bullet$ During  {\em a model training iteration}, DeepMapping trains the weights of the sampled neural architecture in $m\_epochs$ by fixing the controller parameters ($\theta$). We use standard cross entropy~\cite{hinton1995wake} as the loss function to update model weights $M$. Since the sampled neural architecture is a subgraph of the search space, and these layers may be sampled again in future iterations, each model training iteration may improve the accuracy and convergence rate of future model training iterations by sharing parameters across iterations.
Note that each training iteration is run for $\frac{size(D)}{size(D_{batch})}$ training steps.
Since the memorization task may, in practice, need a larger number of iterations to stabilize to a desired level,
usually we choose $N_m  > N_c$, where $N_m$ is the number of model training iterations and $N_c$ is the number of 
controller training iterations.

\subsection{DeepMapping Modification Operations (Desideratum $\mathbf{\#4}$)}\label{sec:data-manipulation}
{

Neural networks are notoriously challenging to be updated:
(a) Incremental training tends to reduce the accuracy of existing data;
(b) It is difficult to tell the neural network to forget something that is already learned, and
(c) Retraining incurs significant latency.
Therefore, to support data modification operations,  \texttt{insert}/\texttt{update}/\texttt{delete}, we piggy-back on the auxiliary structure that we have  described in Section~\ref{sec:accuracy-assure}:

\noindent
\textbf{Insertions.}  
Given a collection of key-value pairs ($D_{insert}$) to be inserted into the DeepMapping's hybrid data structure, we first set the corresponding bit as 1 in the bit vector $V_{exist}$ for the inserted data to indicate these new data exist at query time. Then we run an evaluation of the model on the newly inserted data ($D_{insert}$) to check whether the ($M$) can generalize to them  
by running inference over the keys for the target columns. Only those key-value pairs that are incorrectly inferred need to be inserted into the auxiliary table ($T_{aux}$). 

\begin{algorithm}[h]
\caption{Insert}
\label{alg:insert}
\footnotesize
\begin{algorithmic}[1]
\STATE INPUT: \\
$D_{insert}$: A collection of $n$ tuples to be inserted. \\
$M$: The pre-trained neural network model. \\
$T_{aux}$: The auxiliary table that stores the mis-classified data. \\
$V_{exist}$: The bit
vector for existence check.

\FOR{$i=1 \text{ to } n$}
    \STATE {// \textit{set corresponding existence bit as 1}}
    \STATE $V_{exist}[D_{insert}[i].key] = 1$
    \IF{$f_{decode}(M.infer(D_{insert}[i].key)) = D_{insert}[i].values $}
        \STATE continue
    \ELSE
        \STATE {// \textit{misclassified data is stored in $T_{aux}$}}
        \STATE $T_{aux}.add(D_{insert}[i])$
    \ENDIF
\ENDFOR
\end{algorithmic}
\end{algorithm}

\noindent
\textbf{Deletions.} The deletion process is relatively straightforward as it can be implemented simply by marking the corresponding existence bit as $0$ in the bit vector, $V_{exist}$, to indicate this data does not exist, and delete the ones in the auxiliary table, $T_{aux}$, if it was misclassified, which as detailed in Algorithm ~\ref{alg:delete}. 

\begin{algorithm}[h]
\caption{Delete}
\label{alg:delete}
\footnotesize
\begin{algorithmic}[1]
\STATE INPUT: \\
$D_{delete}$: A collection of $n$ keys to be deleted. \\
$V_{exist}$: A bit vector for existence check.

\FOR{$i=1 \text{ to } n$}
    \STATE {// \textit{set corresponding existence bit as 0}}
    \STATE $V_{exist}[D_{delete}[i]] = 0$
    \IF {$D_{delete}[i]$ in $T_{aux}$}
        \STATE {// \textit{removed the corresponding data from $T_{aux}$, if it exists}}
        \STATE { $T_{aux}.remove(D_{delete}[i])$}
    \ENDIF
\ENDFOR
\end{algorithmic}
\end{algorithm}

\noindent
\textbf{Updates (Substitutions).} We treat updated/replaced key-value pairs ($D_{update}$) as mis-learned data and insert the new values into the auxiliary table ($T_{aux}$) , if they do not have existing entries there. Otherwise, the corresponding entries will be updated in-place.  
Since the keys already exist (otherwise, the process would be an insertion), we do not need to update the existence index. The process is illustrated in Algorithm ~\ref{alg:update}. %

\begin{algorithm}[h]
\caption{Update}
\label{alg:update}
\footnotesize
\begin{algorithmic}[1]
\STATE INPUT: \\
$D_{update}$: a vector of $n$ tuples to be updated. \\
$T_{aux}$: Auxiliary table that stores the mis-classified data.
\FOR{$i=1 \text{ to } n$}
    \IF{$f_{decode}(M.infer(D_{update}[i].key)) = D_{update}[i].values $}
        \STATE $T_{aux}.remove(D_{update}[i])$ 
    \ELSE
        \STATE {// \textit{stored the updates as misclassified data in $T_{aux}$}}
        \IF{$D_{update}[i] \notin T_{aux}$}
            \STATE $T_{aux}.add(D_{update}[i])$ 
        \ELSE 
            \STATE$T_{aux}.update(D_{update}[i])$ 
        \ENDIF
    \ENDIF
\ENDFOR

\end{algorithmic}
\end{algorithm}

DeepMapping retrains the model and reconstructs the auxiliary structures on the underlying data to optimize the compression ratio and query efficiency, only when the auxiliary table becomes too large to satisfy the desiderata in Section~\ref{sec:problem}. Retraining could be performed offline, in background, and/or during non-peak time. In addition, a significant portion of (analytics) database workloads are write-once and read-many~\cite{chen2012interactive, adams2012analysis}, and thus will not frequently trigger retraining.

}

\subsection{Extension to Range Queries}  
DeepMapping can be extended to support range queries using two different approaches. The first approach is based on batch inferences: (1) It applies the range-based filtering over the existence index to collect all keys that fall into the range; (2) It then runs batch inferences on the collected keys to retrieve the corresponding values. The second approach is view-based: (1) It first materializes the results of sampled range queries into a view that contains multiple columns, e.g., range-lower-bound, range-upper-bound, range-query-results; (2) It then learns a DeepMapping structure on top of the materialized view using the range boundaries as the key; (3) At runtime, given the range boundaries, we look up the result in the learned DeepMapping structure. The second approach returns approximate results and is more suitable for range aggregation queries. We will extend DeepMapping to range queries and other types of queries in our future works.

\section{Evaluation}

\subsection{Experimental Environment Setup}
\subsubsection{Workloads} We considered lookup queries and modification queries (\texttt{insert/update/delete}) over three types of datasets with different scale factors (SFs). 

\noindent
\textbf{TPC-H}~\cite{tpch}, with SF $1$ and $10$. We removed attributes that have float-point types, such as quantity and retail\_price, to focus on attributes that have categorical and integer types.

\noindent
\textbf{TPC-DS}~\cite{tpcds}, also with SF $1$ and $10$. Similar to TPC-H, we removed all float-point type attributes. TPC-DS scales domains sub-linearly and avoids the unrealistic table ratios that exist under TPC-H~\cite{poess2002tpc}.

\noindent
\textbf{Synthetic datasets with different levels of key-value correlations
}
We synthesize $1$GB and $10$GB datasets by sampling TPC-H and TPC-DS columns from tables with low/high key-value correlations as explained below. \textit{Each dataset has a key column and single/multiple additional columns.}

The \textit{Single-Column w/ Low Correlation} synthetic dataset is generated by sampling <OrderKey, OrderStatus> mappings from the Order table of TPC-H. In this case, the key-value mapping only exhibits a Pearson correlation of  $1e^{-4}$.%

The \textit{Multiple-Column w/ Low Correlation} synthetic dataset is generated from the TPC-H LineItem table, where the average Pearson correlations of key-value mappings are below $5e^{-4}$. %

The \textit{Single/Multiple-Column w/ High Correlation} datasets are generated from TPC-DS Customer\_Demographics tables. At the same time, the single-column case only selects the CD\_Education\_Status column, and the multiple-column case uses all columns from the table. 
In this table, the key-value mappings show a $0.12$ Pearson correlation on average. The columns from the table exhibit periodical patterns along the key-dimension, which do not exist in the low-correlation dataset.%

\noindent

\textbf{Real-World Dataset} 
We sampled a region of the real-world cropland data from CroplandCROS ~\cite{croplandcros}, an image-based crop distribution data, where each pixel represents one crop type. We preprocessed the data to a three-column tabular data consisting of latitude, longitude, and crop type.

\subsubsection{Hardware Environments}
This experimental study has used three environments: (1) Small-size machine (our target environment): An AWS t2-medium instance with two CPUs and 4 GB memory, which are used to approximate resource-limited edge environments; (2) Medium-size machine: An AWS g4dn.xlarge instance that has 4 CPUs, 16 GB memory, and one T4 GPU with 16 GB memory. All systems run Ubuntu 20.04; (3) Large-size machine: A server that has $24$ CPU cores, $125$ GB memory, and $1$ Nvidia-A10 GPU with $24$ GB memory, which is used to understand DeepMapping's benefits for in-memory computing.

\subsubsection{Comparison Baselines}
We compare the compression ratio and query speed of our DeepMapping (\textbf{DM}) approach against the following baselines:

\begin{itemize}[leftmargin=*]
\item {\em Array-based representations without compression (\textbf{AB}),} where each partition is encoded as a serialized numpy array; 
\item {\em Array-based representations with compression(\textbf{ABC})};
\item {\em Hash-based representations with compression (\textbf{HBC})}; 
\item {\em Hash-based representations without compression (\textbf{HB}),} where each partition is a serialized hash table;
\end{itemize} 

We considered various compression algorithms for DM (i.e., to compress the auxiliary data structure), ABC, and HBC. We used \textbf{ABC-D} to refer to ABC with Dictionary Encoding~\cite{dict-compression}, \textbf{ABC-G} for ABC with Gzip~\cite{gzip}, \textbf{ABC-Z} for ABC with Z-Standard~\cite{rfc8878}, and \textbf{ABC-L} for ABC with LZMA~\cite{lzma}.
For HBC and DM, we only considered thetwo best performing compression algorithms: \textbf{HBC-Z} and \textbf{DM-Z} for HBC and DM with Z-Standard~\cite{rfc8878}, and \textbf{HBC-L} and \textbf{DM-L} for HBC and DM with LZMA~\cite{lzma};
 In addition to the aforementioned lossless compression baselines, we also considered the state-of-the-art semantic compression method, DeepSqueeze~\cite{ilkhechi2020deepsqueeze} (\textbf{DS}). Since it is lossy compression, we set the error bound to $\epsilon = 0.001$. %

\eat{
\begin{figure}[t]

\centering{%
   \includegraphics[width=3.4in]{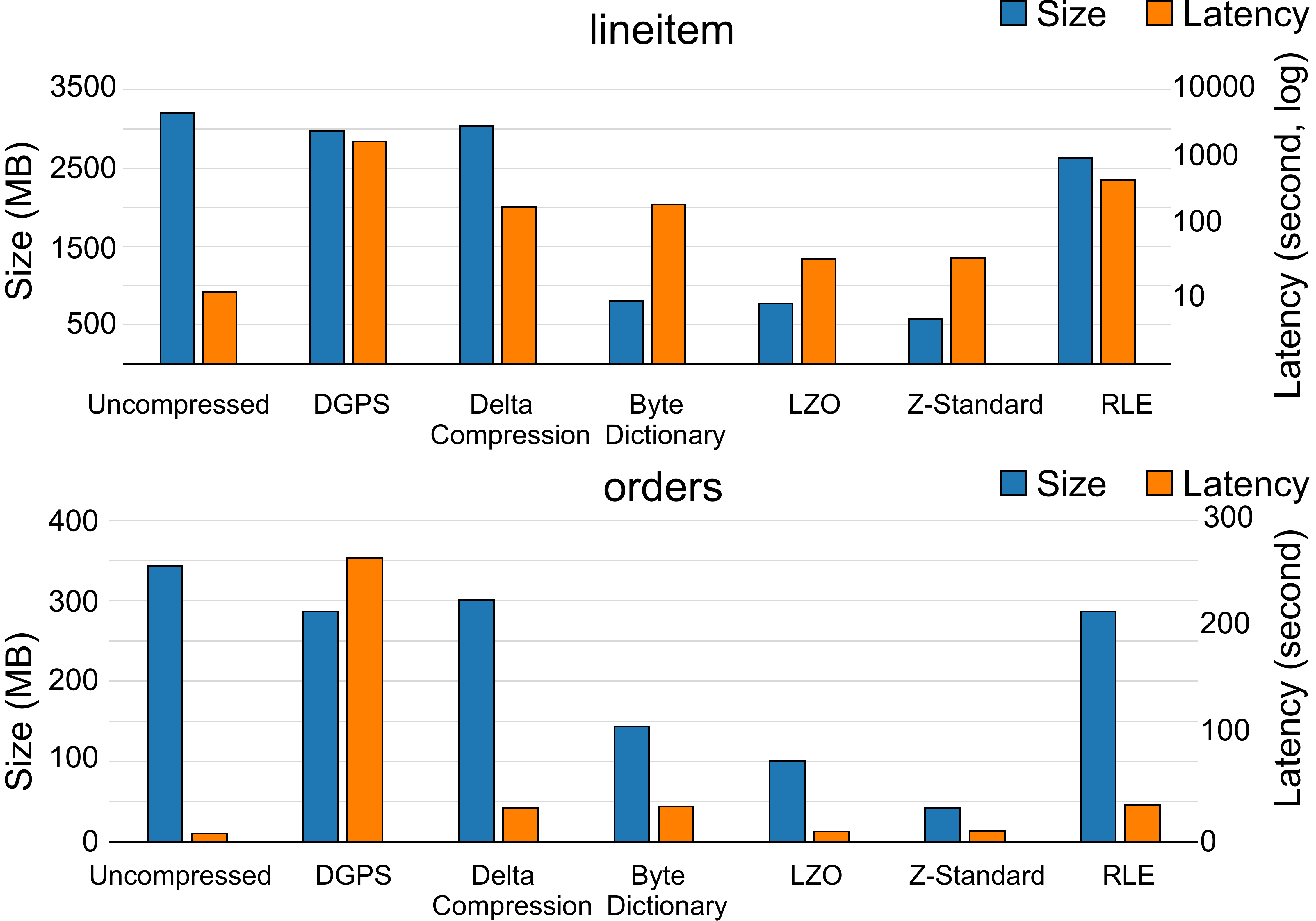}  
}
\caption[]{\label{fig:baseline_compression_comparison} \small Offline storage size and end-end latency on TPC-H, SF=10, B=100,000 with various array-based compression baselines measured in a large machine setup}
\end{figure}
}

 DeepMapping was implemented in Python with the latest numpy, bitarray, and ONNX libraries with C-backends. %
We implemented all baselines in Python with most of the computational intensive functions backed by C/C++, such as dictionary~\cite{python-dict}, binary search,  Numpy functions~\cite{numpy-intro}. 
For Dictionary Encoding, Gzip, LZMA, and Z-Standard, we used public software \cite{gzip-py,lzma-py,lzo-py,zstd-py}, which are all backed by C/C++ with Python bindings. When loading a partition to memory, the corresponding data structure is deserialized. We used the state-of-the-art  Pickle~\cite{pilgrim2009serializing} library for (de)serialization.  
The source code is publically available\footnote{https://github.com/asu-cactus/DeepMapping}. 

\subsubsection{Compression Tuning}

We carefully fine-tuned the compression algorithms for each test case. Taking Z-Standard (ztsd) as an example, %
the default compression level of 1 (fastest) is the best for small-batch queries where the data loading time and the decompression time dominate the end-end latency. %
However, for other cases, after tuning the compression level, zstd can reduce the decompression time up to $30\%$ compared to the default compression level.

\subsubsection{Partition Size Tuning}

For each test case, we use grid search to tune the partition size of DeepMapping and all baselines to ensure the best performance numbers are reported. Our observations are as follows:
(1) For queries over AB and ABC with small batch sizes, the data loading (including deserialization) and decompression time will contribute most to the end-end latency. Therefore, a large partition size, such as $8$MB,  will reduce the frequency of partition loading and decompression and thus reduce latency. However, as the batch size increases, looking up the data in the array starts dominating the overheads, and a small partition size, such as $128$ KB, will benefit large-batch queries. %
(2) For HB and HBC, the lookup complexity is O(1), and the latency is bottlenecked by the deserialization overheads when loading a partition to memory. We observed that deserializing multiple small hash partitions will be faster than deserializing one large partition, and a small partition size, such as 128 KB, leads to optimal performance. 
(3) For DM-Z, similar to AB and ABC, with a small batch size, such as $1000$, the auxiliary data’s loading and decompression time will dominate the latency, and a relatively large partition size, such as 4 MB, can help reduce the latency. However, for large batch sizes, such as $100K$, the latency is less sensitive to the partition size. That is because the auxiliary structure contains only misclassified data and is significantly smaller than the array-based approaches. In the context of DM-L, LZMA introduces additional decompression overhead, which makes the latency sensitive to the partition size. Choosing a small partition size, such as $128$ KB, can enhance query performance.

\subsubsection{Architecture Search }
For MHAS search, we consider a search space of up to two shared hidden layers and two private hidden layers for each task, which serves well for the targeting workloads in this work. For each layer, we also search the number of neurons on the layer in a range of [100, 2000].
As in \cite{pham2018efficient}, the controller is constructed as an LSTM neural network with $64$ hidden units.
During MHAS, the controller parameters, $\theta$, are initialized uniformly in $\mathcal{N}(0, 0.05^2)$ and trained with Adam optimizer at a learning rate of $0.00035$. Each model training iteration uses a learning rate of $0.001$, decayed by a factor of $0.999$.
In each iteration, the model has been trained for $5$ epochs 
{with batch size of $16384$}; the controller is trained for $1$ epoch every $50$ iterations
{with batch size $2048$}. Other configurations are as follows:
(a) number of iterations ($N_t$) for the MHAS model search: $2000$;
(b) number of model training iterations ($N_m$): $2000$;
(c) number of controller training iterations ($N_c$): $40$.
The model search process is followed by training to finetune the accuracy.
Model search and training processes terminate once the absolute value of the changed loss is less than $0.0001$. 

\subsection{Lookup Queries} 

For evaluating the query performance, each time we issue a batch of queries that lookup $B$ randomly selected keys, where $B$ varies from $1,000$ to $100,000$. We test for $5$ times for each experiment and report the average latency.

\subsubsection{Overview} 

We first visualize the trade-offs made by DeepMapping and baselines for TPC-H, TPC-DS, as illustrated in  Figure~\ref{fig:tpch-s10-tradeoff}, and Figure~\ref{fig:tpcds-s10-tradeoff} respectively. 
These results demonstrate that  DeepMapping provides \underline{the best trade-off} among all competitors for an overwhelming majority of the scenarios.
In the TPC-DS benchmark, more columns have large cardinalities (hundreds to thousands of distinct values) than in TPC-H. A consequence is that the memorization is somewhat more complicated, and TPC-DS is generally harder to compress than the TPC-H dataset.
In contrast, some TPC-DS columns strongly correlate with the key column, making these mappings relatively easy to compress. For example, the ${\tt customer\_demographics}$ column achieved a compression ratio of $0.6\%$, reducing the size from $95$MB to $0.5$MB.
The proposed DeepMapping approach can provide significant data size reductions,
with an average of ${\textbf{70.8}}\%$ in TPC-H and TPC-DS (with both SF=1 and 10).%

\eat{
\begin{figure}[t]
\centering{%
   \includegraphics[width=3.4in]{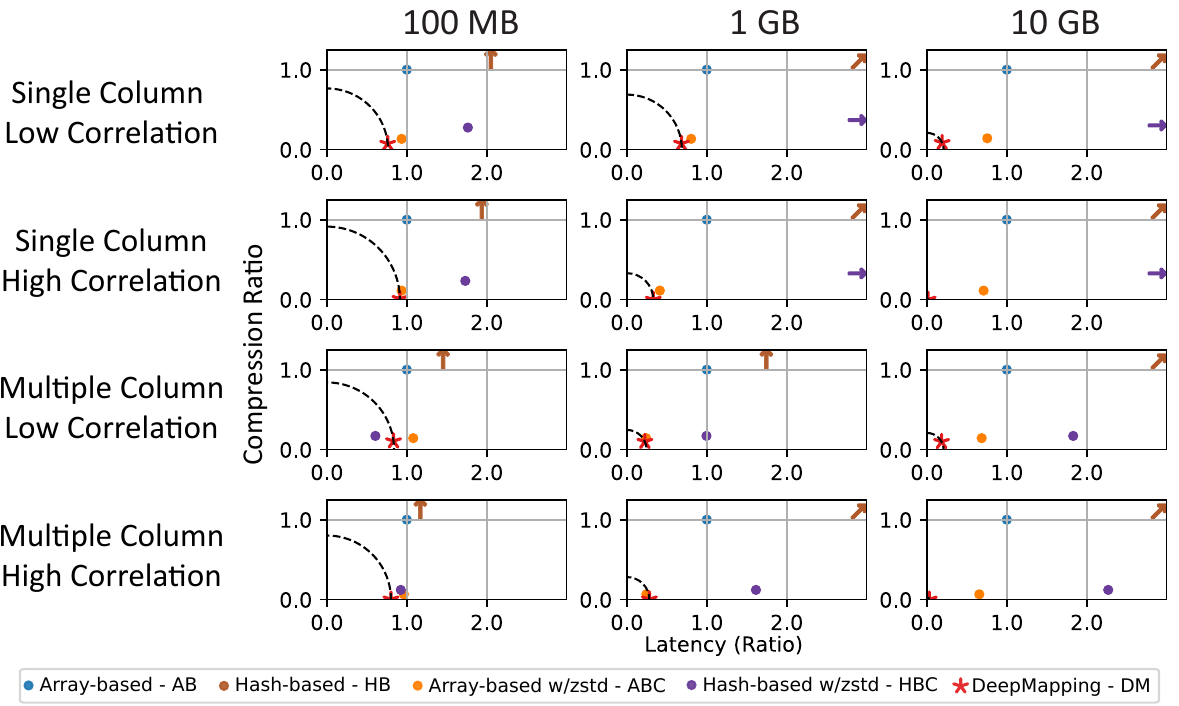}  
}
\caption[]{\label{fig:data-mani-default} \small Trade-off between compression ratio and lookup latency for the synthetic dataset in the small-size machine (B=100,000) -- \textit{Annotations are explained in the footnote $^{\ref{footnote:annotation}}$}}
\end{figure}
}

\begin{figure}[t]
\centering{%
  \includegraphics[width=3.4in]{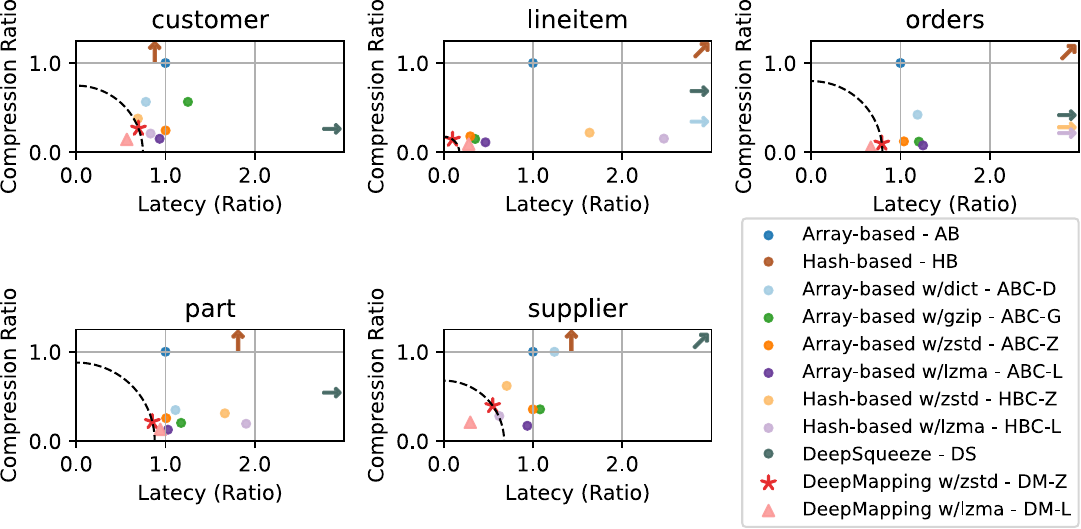}  
}
\caption[]{\label{fig:tpch-s10-tradeoff}\small Trade-off between compression ratio and lookup performance in TPC-H (SF=10, B=100,000) in the small-size machine -- \textit{Annotations are explained in the footnote $^{\ref{footnote:annotation}}$}.
}
\end{figure}

\begin{figure*}[!h]
\centering{%
   \includegraphics[width=0.9\textwidth]{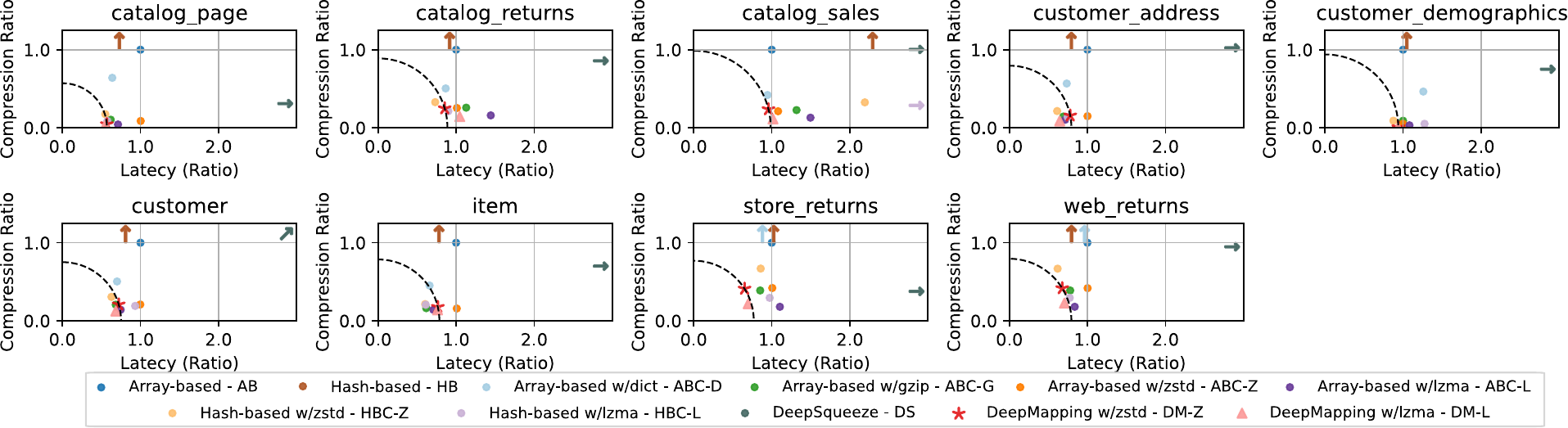}  
}
\caption[]{\label{fig:tpcds-s10-tradeoff}\small Trade-off between compression ratio and lookup performance in TPC-DS (SF=10, B=100,000) in the small-size machine -- \textit{Annotations are explained in the footnote $^{\ref{footnote:annotation}}$. 
}
}
\vspace{-15pt}
\end{figure*}

\footnotetext[2]{{\label{footnote:annotation} In the figures, uncompressed data is always at the point $(1.0,1.0)$). Configurations closer to $(0.0,0.0)$ are more desirable regarding compression ratios and latencies. The dashed arc, $\rightquadcirc$\;\;, indicates the points in the space with the same $L2$ distance to $(0.0,0.0)$ as the DeepMapping algorithm --  hence configurations outside of this arc have a less desirable compression/latency trade-off than DeepMapping; the $\rightarrow$ indicates cases where the access latency is more than $3\times$ slower than accessing the uncompressed data.}}

\subsubsection{Detailed Analysis} 

We first analyze two scenarios: (1) the size of the dataset exceeds memory (Table~\ref{tab:tpch-10-lineitem-small}) vs. (2) the dataset fits into available memory (Table~\ref{tab:tpch-tpcds-small}).

As illustrated in Table~\ref{tab:tpch-10-lineitem-small}, in situations where the dataset exceeds the available memory, the in-memory DeepMapping structure (DM-Z) achieves the best retrieval latency, outperforming other baselines. This is because DeepMapping avoids the decompression overheads and reduces I/O overheads. For example, in TPC-H with SF $10$, the size of the uncompressed Lineitem dataset is about $3.2$ GB in length (after removing the numerical attributes), which is close to the physical memory size in our small-size environment that has $4$ GB memory and exceeds the available memory pool that we set as $3$ GB. As illustrated in Table~\ref{tab:tpch-10-lineitem-small}, %
both the compression ratio and the retrieving speed of DeepMapping (DM-Z and DM-L) significantly outperform the best baselines (i.e., ABC-Z is the fastest and ABC-L achieves the best compression ratio among all baselines except DM-Z and DM-L) up to ${3.4}\times$ and $3.7\times$ respectively. For synthetic datasets, the compression ratio and latency improvements compared to the second best (i.e., ABC-Z for query latency and ABC-L for compression ratio) are up to $43\times$ %
and $44\times$, respectively.%

For the comparisons on the real-world crop dataset, DM-Z outperforms the fastest compression baseline, ABC-Z, by up to $2.08 \times$ in the query lookup. DM-Z also saves $16\%$ more space compared to the most storage-efficient baseline, ABC-L. 
In the rest of the baselines (excluding DM-Z and DM-L), ABC-Z achieves the best query latency, while ABC-L achieves the best compression ratio. However, the LZMA compression algorithm used by ABC-L incurs significantly higher decompression overhead, resulting in slower lookup performance compared to ABC-Z. By applying LZMA on the auxiliary data structure, DM-L can achieve the best storage size among all baselines. %
Compared to ABC-Z, DM-L achieves a speedup of $1.27\times$ in performance and reduces the space by $3\times$. 
DeepSqueeze targets lossy compression. It has to create a set of quantization bins to minimize the errors, which makes it unable to compress the data effectively. Also, the DeepSqueeze relies on a complex AutoEncoder for decompression, leading to high memory consumption and leads to out-of-memory (OOM) errors, while benchmarking datasets that exceed the available memory (e.g., synthetic datasets and the real-world crop dataset on the small-size machine).

\begin{table*}[t]
	\centering
	\scriptsize
	\caption[]{\label{tab:tpch-10-lineitem-small} \small Offline storage size and query latency for datasets that exceed available memory pool on small-size machine.} 
	\begin{tabular}{p{1.8cm}p{2.3cm}p{0.8cm}p{0.8cm}p{0.8cm}p{0.8cm}p{0.8cm}p{0.8cm}p{0.8cm}p{0.8cm}p{0.8cm}p{0.7cm}p{0.7cm}}
		\toprule
		Workloads                                                           & Metrics               & AB     & HB     & ABC-D & ABC-G & ABC-Z & ABC-L & HBC-Z & HBC-L & DS              & DM-Z                & DM-L         \\
		\midrule
		\multirow{ 4}{1.8cm}{TPC-H SF=10 Lineitem}                          & Storage size  (MB)    & 3,203  & 4,698  & 833   & 479   & 573   & 359   & 702   & 491   & 1,928           & 461                 & \textbf{293} \\ 
		                                                                    & Latency, B=1K (sec)   & 1.1    & 2.0    & 1.1   & 1.6   & 1.6   & 2.2   & 1.5   & 1.3   & 32,921          & \textbf{0.3}        & 0.6          \\ 
		                                                                    & Latency, B=10K (sec)  & 5.7    & 13.6   & 9.5   & 7.7   & 7.6   & 9.6   & 10.3  & 12.9  & 38,431          & \textbf{2.0}        & 2.6          \\ 
		                                                                    & Latency, B=100K (sec) & 101    & 297    & 133   & 35    & 30    & 47    & 165   & 249   & 42,779          & \textbf{10}         & 28           \\ 
		\midrule
		\multirow{ 4}{1.8cm}{Synthetic - Single Column w/ Low Correlation}  & Storage size  (MB)    & 10,000 & 23,284 & 4,175 & 1,668 & 1,415 & 815   & 2,084 &  1,089     & 624             & 856                 & \textbf{563} \\
		                                                                    & Latency, B=1K (sec)   & 1.8    & 7.9    & 2.1   & 1.6   & 1.7   & 2.2   & 7.0   &  8.0     & \textit{failed} & \textbf{1.5 }       & 1.6          \\
		                                                                    & Latency, B=10K (sec)  & 22.4   & 92.6   & 23.9  & 20.7  & 19.6  & 25.3  & 67.6  &   74.5    & \textit{failed} & \textbf{15.9}       & 17.2         \\
		                                                                    & Latency, B=100K (sec) & 158    & 742    & 204   & 119   & 119   & 164   & 635   &   820    & \textit{failed} & \textbf{30 }        & 54           \\
		\midrule
		\multirow{ 4}{1.8cm}{Synthetic - Single Column w/ High Correlation} & Storage size (MB)     & 10,000 & 31,691 & 4,171 & 1,563 & 1,127 & 560   & 2,342 &  783     & 648             & \textbf{13}         & \textbf{13}  \\
		                                                                    & Latency, B=1K (sec)   & 2.0    & 16.5   & 4.1   & 1.6   & 1.7   & 2.9   & 15.2  &  20.5     & \textit{failed} & \textbf{0.8 }       & 0.9          \\
		                                                                    & Latency, B=10K (sec)  & 11.8   & 178.2  & 5.1   & 11.2  & 10.1  & 20.6  & 149.4 &  185.4     & \textit{failed} & \textbf{1.2}        & 3.9          \\
		                                                                    & Latency, B=100K (sec) & 148    & 1,112  & 204   & 102   & 105   & 155   & 912   &  1011     & \textit{failed} & \textbf{2}          & 9            \\
		\midrule
		\multirow{ 4}{1.8cm}{Synthetic - Multi Column w/ Low Correlation}   & Storage size  (MB)    & 10,000 & 14,678 & 2,521 & 1,283 & 1,432 & 875   & 1,712 &  1,180     & 2,372           & 984                 & \textbf{678} \\
		                                                                    & Latency, B=1K (sec)   & 1.4    & 6.7    & 4.9   & 1.4   & 1.2   & 2.2   & 4.0   &  4.8     & \textit{failed} & \textbf{1.1 }       & 3.0          \\
		                                                                    & Latency, B=10K (sec)  & 23.9   & 72.1   & 29.6  & 22.3  & 20.5  & 26.3  & 43.6  &  52.9     & \textit{failed} & \textbf{17.8 }      & 18.5         \\
		                                                                    & Latency, B=100K (sec) & 148    & 443    & 153   & 125   & 101   & 141   & 270   &   363    & \textit{failed} & \textbf{27}         & 84           \\
		\midrule
		\multirow{ 4}{1.8cm}{Synthetic - Multi Column w/ High Correlation}  & Storage size  (MB)    & 10,000 & 15,044 & 4,638 & 900   & 666   & 355   & 1,213 &   514    & 2,291            & 26                  & \textbf{21}  \\
		                                                                    & Latency, B=1K (sec)   & 1.8    & 5.6    & 5.2   & 1.3   & 1.4   & 1.8   & 4.4   &   4.5    & \textit{failed} & \textbf{0.3}        & 0.6          \\
		                                                                    & Latency, B=10K (sec)  & 11.8   & 71.3   & 32.4  & 10.4  & 9.1   & 14.9  & 40.0  &   48.5    & \textit{failed} & \textbf{0.9}        & 1.8          \\
		                                                                    & Latency, B=100K (sec) & 145    & 474    & 117   & 191   & 95    & 130   & 329   &   409    & \textit{failed} & \textbf{4}          & 12           \\
            \midrule
		\multirow{ 4}{1.8cm}{Real-world Crop Dataset}  & Storage size  (MB)    & 15,414 & 52,247 & 7,707& 1,672& 1,014      & 341     & 2,599 & 834 & 4,914  & 293  &\textbf{99} \\
		                                                                    & Latency, B=1K (sec)   & 15.7  & 214.6 &5.1 & 7.2& 3.3        & 11.6  & 205.3 & 240 &\textit{failed} &\textbf{0.8} &2.3\\
		                                                                    & Latency, B=10K (sec)  & 33.3 & 342.3 & 9.0& 12.3& 5.9       & 19.9  & 334.5 & 381 &\textit{failed} &\textbf{1.1}&4.1 \\
		                                                                    & Latency, B=100K (sec) & 38.8 & 345.8 & 11.4& 15.0   &9.4     &  24.8  & 332.2 & 386 &\textit{failed} &\textbf{4.5}&7.4 \\
		\bottomrule
	\end{tabular}
\end{table*}

We further compare the performance of DeepMapping and other baselines over datasets that fit the available memory pool size in all three machine environments. Some of the results (for cases with SF=10 and table size larger than $30$MB) are illustrated in Table~\ref{tab:tpch-tpcds-small} highlighting the following findings:

\noindent
 $\bullet$ Even when the dataset fits memory, DeepMapping provides the highest benefits for relatively large datasets, primarily when a strong correlation exists in the underlying data.

\noindent
 $\bullet$ In all considered cases, DeepMapping provides a good compression ratio, easily outperforming the compressed array-based and hash-based representations.

\noindent
 $\bullet$ When the dataset fits into memory, DeepMapping could be slower in query speed than baselines, because the bottleneck is in lookup rather than data-loading.
However, when high key-value correlations exist, DeepMapping spends less time in checking auxiliary table and provides up to $\textbf{1.5}\times$ query speedup (e.g., the customer\_demographics table).

\noindent
 $\bullet$ 
For compressed baselines, the 
need to decompress data will significantly damage the lookup performance compared to the uncompressed baselines in most of cases.

\begin{table*}[t]
	\centering
	\color{black}
	\scriptsize
	\caption[]{\label{tab:tpch-tpcds-small} \small Offline storage size and  query latency for datasets that fit memory pool on small-size (latency-small), medium-size (latency-medium), and large-size (latency-large) machines, B=100,000.} 
	\begin{tabular}{p{1.8cm}p{2.2cm}p{0.8cm}p{0.8cm}p{0.8cm}p{0.8cm}p{0.8cm}p{0.8cm}p{0.8cm}p{0.8cm}p{0.8cm}p{0.8cm}p{0.8cm}p{0.8cm}}
		\toprule
		Workloads                                                  & Metrics              & AB         & HB          & ABC-D      & ABC-G       & ABC-Z & ABC-L      & HBC-Z       & HBC-L & DS    & DM-Z                & DM-L         \\
		\midrule
		\multirow{ 4}{1.8cm}{TPC-H SF=10 Orders}                   & Storage size (MB)    & 343        & 716         & 145        & 41          & 42    & 27         & 67          & 45    & 114   & 34                  & \textbf{19}  \\ 
		                                                           & Latency-Small  (sec) & 8          & 29          & 10         & 9           & 8     & 12         & 31          & 36    & 872   & \textbf{6}          & 7            \\ 
		                                                           & Latency-Medium (sec) & 5          & 12          & 7          & 6           & 5     & 9          & 13          & 16    & 41    & \textbf{3 }         & 5            \\
		                                                           & Latency-Large (sec)  & 4          & 9           & 4          & 5           & 5     & 8          & 10          & 15    & 24    & \textbf{3 }         & \textbf{3}   \\
		\midrule
		\multirow{ 4}{1.8cm}{TPC-H SF=10 Part}                     & Storage size (MB)    & 61         & 110         & 21         & 13          & 17    & \textbf{8} & 19          & 12    & 28    & 13                  & \textbf{8}   \\ 
		                                                           & Latency-Small (sec)  & 8          & 13          & 9          & 13          & 8     & 13         & 13          & 15    & 160   & \textbf{7}          & 8            \\ 
		                                                           & Latency-Medium (sec) & 5          & \textbf{3}  & 6          & \textbf{3 } & 5     & 7          & 13          & 10    & 8     & 4                   & 5            \\
		                                                           & Latency-Large (sec)  & 5          & \textbf{3 } & 4          & 4           & 5     & 6          & \textbf{3 } & 4     & 7     & \textbf{3}          & 4            \\
		\midrule
		\multirow{ 4}{1.8cm}{TPC-DS SF=10 Catalog\_sales }         & Storage size (MB)    & 327        & 681         & 137        & 74          & 69    & 43         & 106         & 68    & 303   & 77                  & \textbf{37}  \\ 
		                                                           & Latency-Small (sec)  & \textbf{8} & 17          & \textbf{8} & 10          & 9     & 12         & 17          & 29    & 1,461 & \textbf{8}          & \textbf{8}   \\ 
		                                                           & Latency-Medium (sec) & 5          & 11          & 7          & 7           & 5     & 8          & 12          & 26    & 55    & \textbf{4 }         & 5            \\
		                                                           & Latency-Large (sec)  & 5          & 9           & 5          & 5           & 5     & 6          & 10          & 20    & 50    & \textbf{4 }         & \textbf{4}   \\
		\midrule
		\multirow{ 4}{1.8cm}{TPC-DS SF=10 Customer\_demographics } & Storage size (MB)    & 95         & 142         & 44         & 9           & 5     & 3          & 9           & 5     & 64    & \textbf{0.5}        & \textbf{0.5} \\ 
		                                                           & Latency-Small (sec)  & 10         & \textbf{9}  & 13         & 10          & 10    & 11         & \textbf{9 } & 13    & 517   & 10                  & 10           \\ 
		                                                           & Latency-Medium (sec) & 4          & 3           & 3          & 5           & 4     & 6          & 3           & 4     & 14    & \textbf{2 }         & \textbf{2}   \\
		                                                           & Latency-Large (sec)  & 6          & 5           & 6          & 5           & 5     & 7          & 6           & 6     & 13    & 6                   & \textbf{4 }  \\
		\midrule
		\multirow{ 4}{1.8cm}{TPC-DS SF=10 Catalog\_returns }       & Storage size (MB)    & 37         & 12          & 19         & 10          & 9     & 6          & 12          & 8     & 29    & 9                   & \textbf{5}   \\ 
		                                                           & Latency-Small  (sec) & 8          & \textbf{6 } & 7          & 9           & 8     & 11         & \textbf{6 } & 7     & 207   & 7                   & 8            \\ 
		                                                           & Latency-Medium (sec) & 6          & 5           & \textbf{4} & 6           & 6     & 8          & \textbf{4 } & 5     & 8     & \textbf{4 }         & 5            \\
		                                                           & Latency-Large  (sec) & 4          & \textbf{3}  & 4          & 4           & 4     & 5          & \textbf{3}  & 4     & 6     & \textbf{3  }        & \textbf{3 }  \\
		\bottomrule
	\end{tabular}
 \vspace{-15pt}
\end{table*}
 
Figure~\ref{fig:tpch-s1-size-breakdown} provides a detailed breakdown of the DeepMapping storage mechanism for TPC-H with different scale factors. As shown in Figure~\ref{fig:tpch-s1-size-breakdown}, the bulk of the storage is taken by the auxiliary table in most scenarios.  Although the model size is significantly smaller than the auxiliary data structure, the model memorized $68\%$ and $66\%$ of the tuples on average for TPC-H SF=1 and SF=10 respectively. 
This observation justifies our MHAS search algorithm, which targets optimizing the entire size of the hybrid architecture rather than searching for a perfect model that achieves $100\%$ accuracy. The observed pattern also applies to other datasets. The models can memorize $78\%$, $80\%$, and $81\%$ of tuples from TPC-DS, SF=10, synthetic datasets, and the real-world cropland dataset, respectively.

\begin{figure}[t]
\centering{%
   \includegraphics[width=3in]{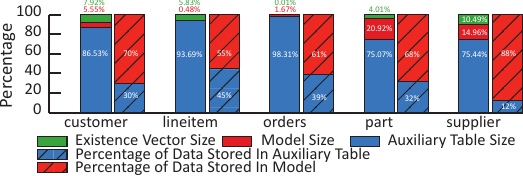}  
}
\centerline{(a) TPC-H (Scale Factor=1)}
\centering{%
   \includegraphics[width=3in]{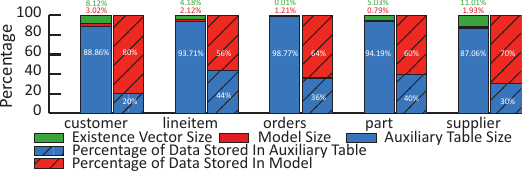}  
}
\centerline{(b) TPC-H (Scale Factor=10)}
\caption{\label{fig:tpch-s1-size-breakdown}\small DeepMapping storage breakdown in the small-size machine.
}
\end{figure}

Figure~\ref{fig:tpch-s1-m100000-perf-breakdown} further shows a breakdown of point query latency over TPC-H tables. For DeepMapping, in most cases, the neural network inference overheads are insignificant, and most of the time is spent querying the auxiliary structure.  DeepMapping significantly reduced the decompression overheads and the I/O overheads, which were included in the purple bar in Figure~\ref{fig:tpch-s1-m100000-perf-breakdown}. %
In addition, we also observed significant runtime memory reduction brought by DeepMapping when the datasets fit into memory, which explained the query latency benefits of DM on the medium and large-size machines as shown in Table~\ref{tab:tpch-tpcds-small}. 

\begin{figure}[t]
 \centering{%
    \includegraphics[width=3.4in]{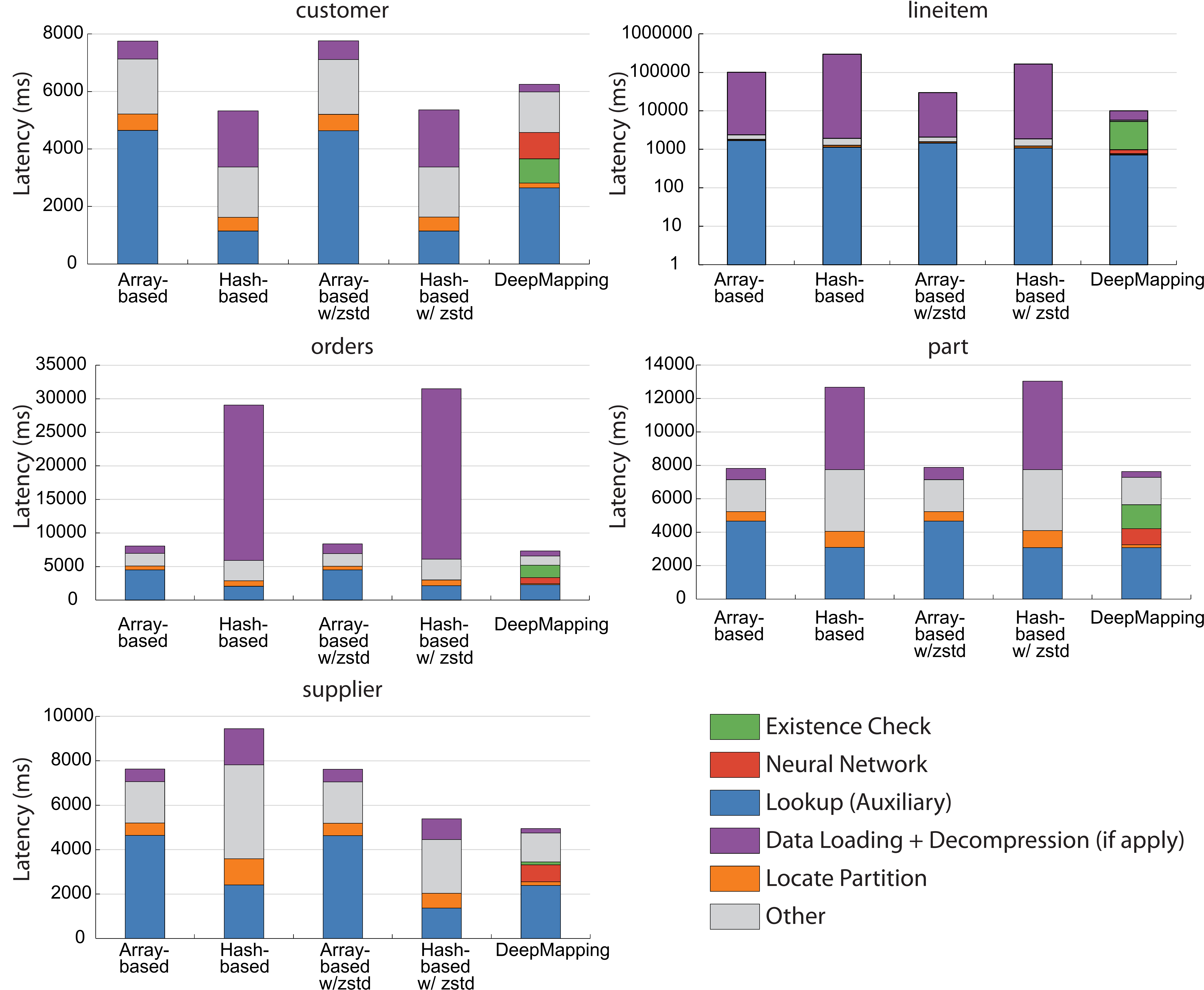}  
 }
 \caption{\label{fig:tpch-s1-m100000-perf-breakdown} \small Breakdown of the end-end latency for TPC-H, SF=10, B=100,000 in the small-size machine. 
 }
 \end{figure}

\subsection{Modification Queries: Insertion/Deletion/Updates} %
\label{sec:modification}

\noindent
\textbf{Insertion.} 
We focus on the synthetic datasets where the value part in the key-value pairs consists of multiple columns. We insert a varying number of key-value pairs following and NOT following the underlying distribution of the data: 
(1) As illustrated in Tab.~\ref{tab:data-manipu-insert-exp}, we first insert low-correlation data (i.e., unseen tuples sampled from the TPC-H Lineitem table) into the low-correlation multi-column synthetic dataset (, which is also sampled from the same table).  Then, we insert high-correlation data into the high-correlation multi-column synthetic dataset.  (2) As illustrated in Tab. ~\ref{tab:data-manipu-insert-not-follow-ori-dist-exp}, we first insert highly-correlated data into the low-correlation synthetic dataset,  and then insert low-correlation data into the high-correlation synthetic dataset.
In Sec.~\ref{sec:modification}, \textbf{DM-Z1} represents DM-Z (DeepMapping using Z-Standard to compress the auxiliary data structure) with retraining triggered after $200MB$ data gets modified, e.g., inserted/deleted, while \textbf{DM-Z} represents DM-Z w/o retraining.)
As illustrated in Tab.~\ref{tab:data-manipu-insert-exp} and Tab. ~\ref{tab:data-manipu-insert-not-follow-ori-dist-exp}, our Deep Mapping approaches DM-Z and DM-Z1 outperformed other baselines for the compressed storage size and the query latency over the compressed data. DM-Z1 can achieve the optimal compression ratio and query speed because DM-Z1 will trigger retraining once. Compared to DM-Z, DM-Z1 leads to a more reasonable hybrid structure with a similar storage size. DM-Z1's searched network consists of a slightly bigger and more accurate model and a smaller auxiliary structure than DM-Z. This structure optimization contributes to the reduction in query latency. As illustrated in Tab. ~\ref{tab:data-manipu-insert-not-follow-ori-dist-exp}, DM-Z trained on the low-correlation synthetic data is robust to the input data that follows the high-correlation distribution.%
However, inserting low-correlation data into high-correlation data leads to a larger auxiliary structure than the results in Tab.~\ref{tab:data-manipu-insert-exp}. The retrained model achieved compression ratios similar to the cases of inserting data of similar distributions in Tab.~\ref{tab:data-manipu-insert-exp}, which demonstrates the model's capability of learning the data mapping. 

\noindent
\textbf{Retraining overheads} is triggered after inserting $200$MB data to the original $1$ GB data ($1.2$GB in total).  For the low-correlation case, the MHAS model search process takes $2$ hours $13$ mins, while the training/fine-tuning after the model search only takes $529$ seconds. The high-correlation case spent $3$ hours $5$ mins in MHAS searching and $694$ seconds in the post-search training. The model search and training time is significantly slower when the datasets exhibit high key-value correlations. That's because the compressibility of the underlying data is higher, and the search space is larger than the low-correlation case. As a result, it takes more time to converge the loss and stop the MHAS and training early.The retraining could run offline, in backgrounds, or at non-peak time.
For DM-Z, which does not retrain, its insertion speed is faster than baselines, as illustrated in Figure~\ref{fig:avg-insertion time}.

In addition, we observed that the query performance of the hashing-based baselines(HB and HBC) is significantly worse than other methods. That is due to the hash table representation, which results in a larger storage size and substantially higher deserialization overheads. Consequently, loading hash partitions from disk to memory is significantly more expensive than other baselines. Therefore, they are significantly slower if many hash partitions cannot fit into and need to be loaded into the memory, as shown in the purple bar in Figure~\ref{fig:tpch-s1-m100000-perf-breakdown}.

\begin{figure}[t]
 \centering{%
    \includegraphics[width=3.4in]{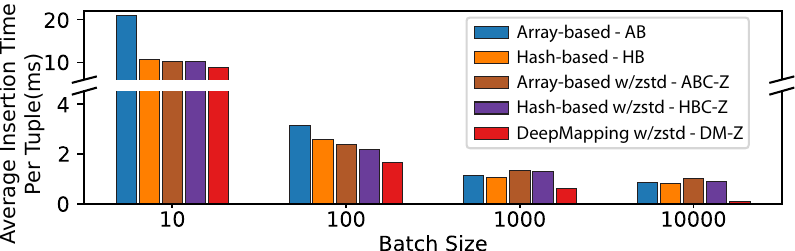}  
 }
 \caption{\label{fig:avg-insertion time} \small Comparison of insertion time with varying batch sizes under the Multi-column with Low Correlation (1GB) %
 }
 \end{figure}

\begin{table}[t]
\centering
\scriptsize
\caption{\label{tab:data-manipu-insert-exp}\small Comparison of the compressed storage size and the query latency (with B = 100,000) after \textbf{inserting} varying data (follows the original distribution) to synthetic datasets using various systems, on the small machine. (DM-Z: w/o retrain; DM-Z1: retrains when 200MB data is inserted)
}
\begin{tabular}{p{2.2cm}p{0.5cm}p{0.5cm}p{0.5cm}p{0.5cm}p{0.5cm}p{0.5cm}p{0.5cm}}
\hline
\toprule
Insertion Size (MB)                 & 0  & 100     & 200     & 300     & 400    & 500     & 600 \\ \hline
\multicolumn{8}{c}{Multi-column with Low Correlation (Uncompressed size: 1GB)}                        \\ \hline
DM-Z-Storage (MB) & \textbf{100}    & \textbf{110}    & \textbf{120}   & \textbf{129}   & \textbf{139}   & \textbf{149}   & \textbf{159}                    \\
DM-Z-Query (ms)      & \textbf{4,483} & \textbf{4,614} & 4,605 & 4,610 & 4,621 & 4,618 & 4,642                  \\ 
\hline
DM-Z1-Storage (MB) &-&-& \textbf{120}                  & \textbf{129} & \textbf{139} & \textbf{149} & \textbf{159} \\
DM-Z1-Query (ms)      &-&-& \textbf{4,302}                   & \textbf{4,306}  & \textbf{4,343}  & \textbf{4,325}  & \textbf{4,330} \\ 
\hline
AB-Storage (MB) & 1,000                                        & 1,100  & 1,200  & 1,300  & 1,400  & 1,500  & 1,600  \\
AB-Query (ms)      & 9,413                                        & 9,215  & 9,307  & 9,858  & 9,933  & 9,515  & 9,343  \\ 
\hline
ABC-Z-Storage (MB) & 143                                          & 158    & 172    & 186    & 200    & 215    & 229    \\
ABC-Z-Query (ms)      & 7,832                                        & 7,539  & 7,269  & 7,380  & 7,178  & 7,304  & 7,381  \\
\hline
HB-Storage (MB) &1,468&1,615&1,762&1,908&2,055&2,202&2,349                 \\ 
HB-Query (ms)      &59,362&59,828&58,116&57,629&59,244&58,246&60,102                 \\ 
\hline
HBC-Z-Storage (MB) &172&189&206&223&240&257&274                    \\
HBC-Z-Query (ms)      &33,977&34,151&33,421&33,933&33,413&33,100&33,595      \\ 
\hline
\hline
\multicolumn{8}{c}{Multi-column with High Correlation (Uncompressed size: 1GB)}              \\ \hline
DM-Z-Storage (MB) & \textbf{5}     & \textbf{14}    & 23    & 33    & 42    & 51    & 61                     \\
DM-Z-Query (ms)     & \textbf{2,210} & \textbf{2,398} & 2,399 & 2,337 & 2,403 & 2,366 & 2,340                  \\
\hline
DM-Z1-Storage (MB) &-&-& \textbf{21}                   & \textbf{30}  & \textbf{40}  & \textbf{49}  & \textbf{58}  \\
DM-Z1-Query (ms)      &-&-& \textbf{2,197}                   & \textbf{2,188}  & \textbf{2,293}  & \textbf{2,292}  & \textbf{2,285}  \\
\hline
AB-Storage (MB) & 1,000                                        & 1,100  & 1,200  & 1,300  & 1,400  & 1,500  & 1,600  \\
AB-Query (ms)      & 9,998                                        & 9,543  & 9,517  & 9,972  & 9,933  & 9,741  & 10,122 \\ 
\hline
ABC-Z-Storage (MB) & 67                                           & 73     & 80     & 87     & 93     & 100    & 107    \\
ABC-Z-Query (ms)      & 6,823                                        & 6,867  & 6,875  & 6,759  & 7,077  & 6,999  & 6,814  \\
\hline
HB-Storage (MB) &1,504&1,655&1,805&1,956&2,106&2,256&2,407                 \\ 
HB-Query (ms)      &63,330&62,595&61,269&62,445&65,629&61,940&61,877                 \\ 
\hline
HBC-Z-Storage (MB) &121&133&146&158&170&182&194                 \\
HBC-Z-Query (ms)      &32,416&32,118&32,142&32,293&33,085&33,668&32,220  \\           
\hline
\bottomrule
\end{tabular}
\end{table}

{
\color{blue}
\begin{table}[t]
\centering
\scriptsize
\caption[]{\label{tab:data-manipu-insert-not-follow-ori-dist-exp}\small Comparison of the compressed storage size and the query latency (with B = 100K) after \textbf{inserting} varying data that \textbf{does not} follow the original distribution on the small machine. (DM-Z: w/o retrain; DM-Z1: retrains when 200MB data is inserted)
}
\begin{tabular}{p{2.2cm}p{0.5cm}p{0.5cm}p{0.5cm}p{0.5cm}p{0.5cm}p{0.5cm}p{0.5cm}}
\hline
\toprule
Insertion Size (MB)                 & 0  & 100     & 200     & 300     & 400    & 500     & 600 \\ \hline

\multicolumn{8}{c}{Multi-column with Low Correlation (Uncompressed size: 1GB)}                        \\ \hline
DM-Z-Storage (MB) & \textbf{100}    & \textbf{109}    & 119   & 129  & 139  & 149   & 158               \\
DM-Z-Query (ms)      & \textbf{4,483} & \textbf{4,468} & 4,691 & 4,588 & 4,557 & 4,637 & 4,556                  \\ 
\hline
DM-Z1-Storage (MB) &-&-& \textbf{104}                  & \textbf{105} & \textbf{107} & \textbf{109} & \textbf{111} \\
DM-Z1-Query (ms)      &-&-& \textbf{4183}                   & \textbf{4,173}  & \textbf{4,253}  & \textbf{4,184}  & \textbf{4,180} \\ 
\hline
AB-Storage (MB) & 1,000                                        & 1,100  & 1,200  & 1,300  & 1,400  & 1,500  & 1,600  \\
AB-Query (ms)      & 9,413                                        & 9,457  & 9,564  & 9,9351  & 9,501  & 9,151  & 9,862  \\ 
\hline
ABC-Z-Storage (MB) & 143                                          & 146    & 149    & 151    & 154    & 157    & 159    \\
ABC-Z-Query (ms)      & 7,832                                        & 7,978  & 8,012  & 7,703  & 7,679  & 8,129  & 8,155  \\
\hline
HB-Storage (MB) &1,468&1,615&1,762&1,908&2,055&2,202&2,349                 \\ 
HB-Query (ms)      &59,362&59,519&60,795&59,339&60,337&60,712&61,015                 \\ 
\hline
HBC-Z-Storage (MB) &172&177&183&189&195&200&206                    \\
HBC-Z-Query (ms)      &33,977&29,344&29,407&29,458&28,839&29,466&30,992      \\ 
\hline
\hline
\multicolumn{8}{c}{Multi-column with High Correlation (Uncompressed size: 1GB)}              \\ \hline
DM-Z-Storage (MB) & \textbf{5}     & \textbf{18}    & 31    & 44    & 57    & 70    & 83                     \\
DM-Z-Query (ms)     & \textbf{2,210} & \textbf{2253} & 2245 & 2231 & 2256 & 2274 & 2237                  \\
\hline
DM-Z1-Storage (MB) &-&-& \textbf{23}                   & \textbf{32}  & \textbf{41}  & \textbf{50}  & \textbf{59}  \\
DM-Z1-Query (ms)      &-&-& \textbf{2,173}                   & \textbf{2,189}  & \textbf{2,207}  & \textbf{2,167}  & \textbf{2,177}  \\
\hline
AB-Storage (MB) & 1,000                                        & 1,100  & 1,200  & 1,300  & 1,400  & 1,500  & 1,600  \\
AB-Query (ms)      & 9,998                                        & 9,921  & 9,676  & 9,512  & 9,609  & 9,851  & 9,743 \\ 
\hline
ABC-Z-Storage (MB) & 67                                           & 83     & 99     & 116     & 132     & 148    & 164    \\
ABC-Z-Query (ms)      & 6,823                                        & 6,905  & 6,922  & 6,878  & 6,613  & 6,840  & 6,826  \\
\hline
HB-Storage (MB) &1,504&1,655&1,805&1,956&2,106&2,256&2,407                 \\ 
HB-Query (ms)      &63,330&63,796&64,831&61,905&62,014&60,647&61,456                 \\ 
\hline
HBC-Z-Storage (MB) &121&141&160&180&199&218&238                 \\
HBC-Z-Query (ms)      &32,416&34,160&35,959&34,647&36,736&34,618&35,713  \\           
\hline
\bottomrule
\end{tabular}
\end{table}
}

\eat{
\begin{table}[t]
\centering
\scriptsize
\caption{\label{tab:??}\small Model Searching Time and Training Time of when training is triggered.}
\begin{tabular}{lc}
\toprule
 \multicolumn{2}{l}{Multiple Value Columns with Low Correlation (Uncompressed size: 1GB)}  \\ \hline
 Model Searching Time                              &                                       \\
 Model Training Time                               & 529 seconds                           \\ \hline
\multicolumn{2}{l}{Multiple Value Columns with High Correlation (Uncompressed size: 1GB)} \\ \hline
Model Searching Time                              &                                       \\
Model Training Time                               & 694 seconds     \\           \bottomrule
\end{tabular}
\end{table}
}

\eat{

}

\noindent
\textbf{Deletion and Updates.} As illustrated in Table~\ref{tab:data-manipu-delete-exp}, DM and DM1 outperform the best of baselines by $1.5\times$ to more than $10\times$ in query speed and $1.3\times$ to more than $10\times$ in compression ratio. DM1's query speed is slower than DM in a few cases because of the randomness in decompressing $V_{exist}$ (due to the random distribution of the $1$ bits). Other observations are similar to the insertion case. The update operation is similar to an insertion. Therefore, we do not elaborate on evaluation results for the update operation to save space. 

\begin{table}[t]
\centering
\scriptsize
\caption{\label{tab:data-manipu-delete-exp}\small Comparison of the compressed storage size and the query latency (with B = 100,000) after \textbf{deleting} varying data from synthetic datasets using various systems, on the small machine. (DM-Z: w/o retrain; DM-Z1: retrains when 200MB data is deleted)
}
\begin{tabular}{p{2.2cm}p{0.5cm}p{0.5cm}p{0.5cm}p{0.5cm}p{0.5cm}p{0.5cm}p{0.5cm}}
\hline
\toprule
Deletion Size (MB)                 & 0  & -100     & -200     & -300     & -400    & -500     & -600 \\ \hline
\multicolumn{8}{c}{Multi-column with Low Correlation (Uncompressed size: 1GB)}  \\ \hline
DM-Z-Storage (MB)        & \textbf{100}    & \textbf{92}    & \textbf{84}    & \textbf{75}    & \textbf{66}   & \textbf{56}   & \textbf{47}   \\
DM-Z-Query (ms)          & \textbf{4,483}    & \textbf{3,885}    & 3,824    & 4,000    & 3,779   & 3,834   & 3,771   \\ \hline
DM-Z1-Storage (MB)       &- &  -& \textbf{84}    & \textbf{75}    & \textbf{66}   & \textbf{56}   & \textbf{47}   \\
DM-Z1-Query (ms)          & - & -& \textbf{3,496}    & \textbf{3,430}    & \textbf{3,310}   & \textbf{3,321}   & \textbf{3,169}   \\ \hline
AB-Storage (MB)        & 1000     & 900      & 800      & 700      & 600     & 500     & 400     \\
AB-Query (ms)          & 9,413    & 7,548    & 7,998    & 7,064    & 6,402   & 6,207   & 6,267   \\ \hline
ABC-Z-Storage (MB)        & 143   & 129  & 115   & 102   & 88   & 74   & 60   \\
ABC-Z-Query (ms)          & 7,794    & 6,428    & 6,470    & 6,101    & 5,841   & 5,719   & 5,444   \\ \hline
HB-Storage (MB)        & 1,504       & 1,322       & 1,175       & 1,029       & 882         & 736      & 589      \\
HB-Query (ms)          & 63,330       & 53,944       & 49,635       & 42,387       & 28,444      & 19,538      & 20,381      \\ \hline
HBC-Z-Storage (MB)        & 172       & 156       & 140       & 124       & 108      & 91      & 75      \\
HBC-Z-Query (ms)          & 33,977       & 24,426       & 21,672       & 19,498       & 17,730      & 14,715      & 12,910      
\\ 
\hline
\hline
\multicolumn{8}{c}{Multi-column with High Correlation (Uncompressed size: 1GB)} \\ \hline
DM-Z-Storage (MB)      & \textbf{5}        & \textbf{6}        & 6        & 6        & 5        & 5        & \textbf{4}       \\
DM-Z-Query (ms)        & \textbf{2,210}    & \textbf{2,180}    & \textbf{2,038}    & \textbf{1,933}    & \textbf{1,844}    & \textbf{1,818}    & 1,753   \\ \hline
DM-Z1-Storage (MB)      & -        & -        & \textbf{4}        & \textbf{4}        & \textbf{4}        & \textbf{4}        & \textbf{4}       \\
DM-Z1-Query (ms)        & -        & -        & 2,211    & 2,092    & 1,954    & 1,839    & \textbf{1,676}   \\ \hline
AB-Storage (MB)      & 1000     & 900      & 800      & 700      & 600      & 500      & 400     \\
AB-Query (ms)        & 9,998    & 8,295    & 8,136    & 8,148    & 7,150    & 6,657    & 6,054   \\ \hline
ABC-Z-Storage (MB)      & 67       & 64       & 58       & 53       & 47       & 41       & 35      \\
ABC-Z-Query (ms)        & 6,823    & 6,660    & 6,172    & 6,551    & 6,018    & 5,982    & 5,931   \\ \hline
HB-Storage (MB)      & 1,504    & 1,354    & 1,204    & 1,054    & 904      & 754      & 604     \\
HB-Query (ms)        & 63,330   & 56,836   & 52,929   & 48,052   & 35,353   & 20,715   & 20,862  \\ \hline
HBC-Z-Storage (MB)      & 121      & 112      & 102      & 90       & 78       & 66       & 55      \\
HBC-Z-Query (ms)        & 32,416   & 30,987   & 23,187   & 20,906   & 18,723   & 15,900   & 13,721  \\ \hline
\bottomrule
\end{tabular}
\end{table}

\subsection{Multi-Task Hybrid Architecture Search (MHAS)} 
We have discussed the MHAS overheads (i.e., retraining) in Sec.~\ref{sec:modification}
In this section, we further evaluate the effectiveness of our proposed MHAS algorithm, using the TPC-H, SF=1. %
In Figure~\ref{fig:nas-tpch-s1-overall}, we plot the 
sampled models' compression ratio against the controller's training process. 

\begin{figure}[h]
\centering{%
    \vspace{-10pt}
   \includegraphics[width=3.2in]{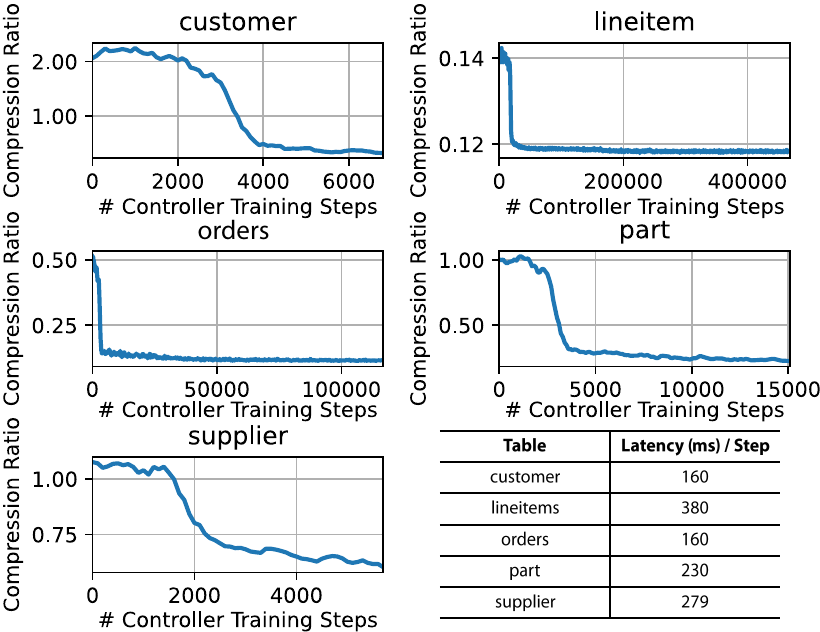}  
}
\caption{\label{fig:nas-tpch-s1-overall}\small Compression ratios during MHAS (TPC-H, scale =1; plots  smoothed with running average window of $500$).}
\vspace{-5pt}
\end{figure}

\noindent

\textbf{Training Time vs. Compression Time} 
Taking the LineItem table from TPC-H SF=1, as an example, DM takes 1 hour and 25 mins for the model search (MHAS) to converge and takes 5 mins for model training. DeepSqueeze (DS) requires 11 mins to encode the data. Z-Standard takes 80 seconds for compression. LZMA spends 86 seconds in compression. In addition, Hashtable with Z-Standard (HBC-Z) and LZMA (HBC-L) spend 82 and 152 seconds in compression respectively.

Despite DeepMapping's significant improvements in balancing compression ratio and query performance, the search and training of the neural network model is expensive. Our future work will optimize this process for DeepMapping, e.g., leveraging model reuse and transfer learning~\cite{zhou2022benchmark}.

As we see here, at the very beginning of the search stage, there is a "flat" region where the compression ratio is not yet decreasing -- this is because, at the early stages, the sampled models are not yet capable of memorizing the data.
In fact, at this stage, the data structure size may be larger than the original data since much of the memorization work is left to the auxiliary table.
As the controller training proceeds, however, the sampled models are quickly getting better at memorizing the data, and the compression ratio improves significantly.

Figure~\ref{fig:nas-tpch-s1-part-sampled-models} illustrates the trade-ff between compression ratio and latency during the search. In the figure, each dot corresponds to a sampled architecture, and each color corresponds to a certain search stage. Initially, samples may cover a large range, indicating that the model search has not stabilized. However, as the search progresses, the samples start clustering in an increasingly shrinking region in the search space, illustrating the effectiveness of the MHAS strategy.

\begin{figure}[h]
\centering{%
   \includegraphics[width=3.2in]{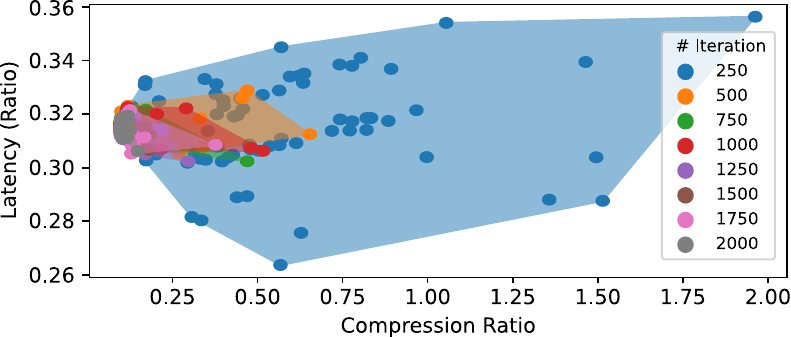}  
}
\caption{\label{fig:nas-tpch-s1-part-sampled-models} \small
Progression of compression ratio vs. latency trade-off during MHAS search process (TPC-H ${\tt part}$ table, scale = 1; each dot corresponds to a sampled architecture)%
}
\end{figure}

\vspace{-10pt}
\section{Conclusions}
\label{findings}

In this work, we proposed a novel DeepMapping structure to achieve lossless compression and desirable query speed at the same time. It is achieved by automatically searching multi-tasking deep neural networks to model key-value mappings in relational tables while using an auxiliary structure to manage misclassified data and support \texttt{insert/update/delete}. The evaluation observations include:

\noindent
\textbf{$\bullet$} DeepMapping achieves the best compression ratio and retrieval speeds for large datasets, especially when the key strongly correlates to the value. When the uncompressed dataset exceeds memory but the DeepMapping structure fits into memory (ensured by the MHAS), DeepMapping helps significantly reduce the I/O and decompression overheads and outperforms other baselines.

\noindent
\textbf{$\bullet$} The hashing baseline may achieve better retrieval speeds for small datasets that fit memory because the hashtable lookup operation is cheaper than inference computations. However, in these cases, DeepMapping achieves a better compression ratio than the hashing baseline applied with state-of-the-art compression techniques. For large-scale datasets that do not fit into memory, hashing-based approaches perform the worst due to the deserialization complexity of hash structures.

\noindent
\textbf{$\bullet$} The MHAS framework can effectively decide whether a proper DeepMapping structure exists for the given tabular dataset that can gain compression size and retrieval speeds.

\eat{
\vspace{3pt}
In the future, we will explore several open questions:

\noindent
\textbf{$\bullet$} How to integrate DeepMapping with error-bounded numerical data compression and retrieving approaches?

\noindent
\textbf{$\bullet$} How to support more complicated query workloads, such as aggregation and join? DeepMapping can be easily used to implement hash aggregation and hash join. However, searching for a model that accurately predicts one-to-many \texttt{groupBy} and \texttt{join} relationships is a challenging problem.

\noindent
\textbf{$\bullet$} How to extend DeepMapping to compress large-scale datasets (e.g., terabyte-level) in the cloud environment?
}

\section{Acknowledgment}

We thank all anonymous reviewers for their valuable feedback. Special thanks to Rajan Hari Ambrish for insightful discussion and Pratanu Mandal, for providing essential crop dataset information. This work is supported by the Amazon Research Award, NSF grants \#2144923 and \#2311716, U.S. Department of Homeland Security under Grant Award Number 17STQAC00001-07-00, and U.S. USACE GR40695.

\section{Disclaimer}

The views and conclusions contained in this document are those of the authors and should not be interpreted as necessarily representing the official policies, either expressed or implied, of the U.S. Department of Homeland Security.

\bibliographystyle{IEEEtran}
\bibliography{refs}

\end{document}